\documentclass[preprint,aps,amsmath,nofootinbib,tightenlines,superscriptaddress]{revtex4}
\usepackage{graphicx}
\usepackage{amsmath}
\usepackage{slashed}
\usepackage{bm}
\usepackage{epsfig}
\usepackage[dvips]{hyperref}

\newcommand{\cn}{\overline{\xi}_{n}}
\newcommand{\cnW}{\bar{ \chi}}

\newcommand{\hv}{h_{v}}

\newcommand{\pbd}{\overline{\mathcal{P}}^{\dagger}} %P-bar dagger
\newcommand{\pb}{\overline{\mathcal{P}}}%P-bar
\newcommand{\pp}{\mathcal{P}^{\bot}}%P-perp
\newcommand{\ppd}{\mathcal{P}^{\dagger}_{\! \bot}}%P-perp-dagger
\newcommand{\ppda}{\mathcal{P}^{\bot \dagger}_{\alpha}}%P-perp-dagger-\alpha
\newcommand{\ppdb}{\mathcal{P}^{\bot \dagger}_{\beta}}%P-perp-dagger-\beta
\newcommand{\ppds}{\slashed{\mathcal{P}}_{\! \bot}^{\dagger}}%P-perp-dagger-slash

\newcommand{\nb}{\bar{n}}%nbar
\newcommand{\nbs}{\slashed{\nb}}%n-bar-slash
\newcommand{\ns}{\slashed{n}}%n-bar-slash
\newcommand{\nbdv}{\nb \! \cdot \! v}%nbar dot v
\newcommand{\ndv}{n \! \cdot \! v}
\newcommand{\Nbs}{\frac{\nbs}{\nbdv}}
\newcommand{\Nm}{\frac{n^{\mu}}{n \! \cdot \! v}}

\newcommand{\gpab}{g^{\alpha \beta}_{\bot}}%g perp \alpha \beta
\newcommand{\gpam}{g^{\alpha \mu}_{\bot}}%g perp \alpha \mu
\newcommand{\cB}{\mathcal{B}}%curly B
\newcommand{\cBp}{\mathcal{B}_{\! \bot}}%curly B perp
\newcommand{\cBps}{\slashed{\mathcal{B}}_{\! \bot}}%curly B perp slash
\newcommand{\cBpa}{\mathcal{B}^{\bot}_{\alpha}}%curly B perp alpha
\newcommand{\cBpb}{\mathcal{B}^{\bot}_{\beta}}%curly B perp beta

\newcommand{\deltas}[1]{\delta^{(\lambda^{#1})}}%the s stands for scet
\newcommand{\Deltap}{\Delta_{\! \bot}}%Delta perp
\newcommand{\Deltapa}{\Delta^{\! \bot}_{\alpha}}%Delta perp alpha
\newcommand{\Deltapb}{\Delta^{\! \bot}_{\beta}}%Delta perp \beta
\newcommand{\Deltaps}{\slashed{\Delta}_{\! \bot}}%Delta perp slash
\newcommand{\deltaa}{\delta^{\alpha}}%see eqn (10)

\newcommand{\gammapa}{\gamma^{\alpha}_{\bot}}%gamma perp alpha
\newcommand{\gammapb}{\gamma^{\beta}_{\bot}}%gamma perp \beta

\newcommand{\sub}[1]{_{(#1)}}%for labeling currents and dirac structures etc. i didnt start using this right away

\newcommand{\cP}{{\mathcal P}}

\renewcommand{\O}{\ensuremath{\mathcal{O}}}

\newcommand{\dd}[1]{\mathop{\mathrm{d}{#1}}}%differential d
\newcommand{\pleft}{\ensuremath{\overleftarrow{\partial}}}

\newcommand{\cDgppPl}{\,\overleftarrow {\mathcal D}{}_{\!c\perp}^{\alpha}}

\newcommand{\cDpusLeft}[1]{\,i\!\overleftarrow {\mathcal D}_{us}^{\perp\, #1}}
\newcommand{\cDpusRight}[1]{\,i\!\overrightarrow {\mathcal D}_{us}^{T\, #1}}

\newcommand{\mcdot}{\!\cdot\!}
\newcommand{\bn}{{\bar n}}
\newcommand{\bnP}{\bar {\mathcal P}}

\newcommand{\nslash}{n\!\!\!\slash}
\newcommand{\bnslash}{\bar n\!\!\!\slash}

\renewcommand{\L}{\ensuremath{\mathcal{L}}}
\newcommand{\eq}{\ensuremath{\mspace{-5mu}=\mspace{-3mu}}}
\newcommand{\plus}{\ensuremath{\!+\!}}
\newcommand{\minus}{\ensuremath{\!-\!}}

\newcommand{\ot}{\ensuremath{\otimes}}

\newcommand{\tr}[1]{\ensuremath{\mathop{\mathrm{tr}}{#1}}}

\newcommand{\half}{\ensuremath\frac{1}{2}}

\newcommand{\hc}[1]{\ensuremath{{#1}^{\dagger}}}
\newcommand{\ffrac}[2]{\frac{#1}{#2}} 
\renewcommand{\r}[1]{\ensuremath{\mathrm{#1}}}

\newcommand{\beq}{\begin{equation}}
\newcommand{\eeq}{\end{equation}}
\newcommand{\bal}{\begin{align}}
\newcommand{\eal}{\end{align}}
\newcommand{\bsp}{\begin{split}}
\newcommand{\esp}{\end{split}}
\newcommand{\ben}{\begin{enumerate}}
\newcommand{\een}{\end{enumerate}}
\newcommand{\bca}{\begin{cases}}
\newcommand{\eca}{\end{cases}}
\newcommand{\bpm}{\begin{pmatrix}}
\newcommand{\epm}{\end{pmatrix}}
\newcommand{\bbm}{\begin{bmatrix}}
\newcommand{\ebm}{\end{bmatrix}}
\newcommand{\bsm}{\begin{smallmatrix}}
\newcommand{\esm}{\end{smallmatrix}}
\newcommand{\bvm}{\begin{vmatrix}}
\newcommand{\evm}{\end{vmatrix}}
\newcommand{\bmx}[1]{\left(\begin{array}{*{#1}{c}}}
\newcommand{\emx}{\end{array}\right)}
\newcommand{\bbmx}[2]{\renewcommand{\arraystretch}{#2}\left[\begin{array}{*{#1}{c}}}
\newcommand{\ebmx}{\end{array}\right]}
\newcommand{\bpmx}[2]{\renewcommand{\arraystretch}{#2}\left(\begin{array}{*{#1}{c}}}
\newcommand{\epmx}{\end{array}\right)}
\newcommand{\bmxw}[1]{\renewcommand{\arraystretch}{2}\left(\begin{array}{*{#1}{c}}}
\newcommand{\bmxww}[1]{\renewcommand{\arraystretch}{2.5}\left(\begin{array}{*{#1}{c}}}
\newcommand{\bdet}[1]{\renewcommand{\arraystretch}{2.5}
    \left|\begin{array}{*{#1}{c}}}
\newcommand{\bndet}[1]{\renewcommand{\arraystretch}{1.5}
      \left|\begin{array}{*{#1}{c}}}
\newcommand{\edet}{\end{array}\right|\renewcommand{\arraystretch}{1}}
\newcommand{\bea}{\begin{eqnarray}}
\newcommand{\eea}{\end{eqnarray}}
\newcommand{\bit}{\begin{itemize}}
\newcommand{\eit}{\end{itemize}}

\renewcommand{\l}[1]{\label{#1}}
\newcommand{\Eqr}[1]{Eq. (\ref{#1})}

\newcommand{\nn}{\nonumber}

\newcommand{\fbrac}[2]{\frac{#1}{(#2)}}

\newcommand{\SCETa}{\ensuremath{\textrm{SCET}_{\textrm{I}}\ }}
\newcommand{\OMIT}[1]{}

\begin{document}
\setlength\baselineskip{15pt}

\preprint{ \vbox{ 
\hbox{hep-ph/0508214} 
\hbox{MIT-CTP-3667} 
\hbox{DOE/ER/40762-347}
} }
%\preprint{hep-ph/yymmxxx}
%\preprint{MIT-CTP-3667}
%\preprint{DOE/ER/40762-347}

\title{\phantom{x}\vspace{0.5cm}
Constraint equations for heavy-to-light currents in SCET
\vspace{0.6cm}
}

\author{Christian M.  Arnesen}
\email{arnesen@mit.edu}
\affiliation{Center for Theoretical Physics, Massachusetts
Institute of Technology,\\ Cambridge, MA 02139, USA
\vspace{0.2cm}}

\author{Joydip Kundu}
\email{jkundu@physics.umd.edu}
\affiliation{Department of Physics,
University of Maryland, College Park, MD 20742, USA
\vspace{0.4cm}}

\author{Iain W. Stewart\vspace{0.5cm}}
\email{iains@mit.edu}
\affiliation{Center for Theoretical Physics, Massachusetts
Institute of Technology,\\ Cambridge, MA 02139, USA
\vspace{0.2cm}}

%\date{\today}
%\vspace{0.5cm}

\begin{abstract}
\vspace{0.3cm}

A complete basis for the next-to-next-to leading order heavy-to-light currents
in the soft-collinear effective theory is constructed.  Reparameterization
invariance is imposed by deriving constraint equations. Their solutions give the
set of allowed Dirac structures as well as relations between the Wilson
coefficients of operators that appear at different orders in the power
expansion. The completeness of reparameterization invariance constraints derived
on a projected surface is investigated.  We also discuss the universality of the
ultrasoft Wilson line with boundary conditions.

\end{abstract}

\maketitle

%%%%%%%%%%%%%%%%%%%%%%%%%%%%%%%%%%%%%%%%%%%%%%%%%%%%%%%%%%%%%%%%%%%%%%%%
\section{Introduction}
%%%%%%%%%%%%%%%%%%%%%%%%%%%%%%%%%%%%%%%%%%%%%%%%%%%%%%%%%%%%%%%%%%%%%%%%

The soft-collinear effective theory (SCET) provides a systematic approach for
separating hard, soft, and collinear dynamics in processes with energetic quarks
and gluons~\cite{Bauer:2000ew,Bauer:2000yr,Bauer:2001ct,Bauer:2001yt}. In SCET
the infrared physics is described by effective theory fields with well defined
momentum scaling, which are used to build operators order by order in the power
expansion. The hard perturbative corrections are contained in the Wilson
coefficients which can be computed by matching computations order by order in
perturbation theory in $\alpha_s$. The symmetries and power counting in the
effective theory simplify the derivation of factorization theorems and provide a
systematic method of treating power suppressed contributions.  The construction
of the complete set of allowed operators for a process is one of the first steps
towards deriving factorization theorems. The operators are constrained by gauge
symmetry in the effective theory, as well as by heavy quark effective theory
(HQET) and SCET reparameterization invariance
(RPI)~\cite{Luke:1992cs,Chay:2002vy,Manohar:2002fd}.  The operators and Wilson
coefficients are typically coupled by a convolution integral over the large
momenta of gauge invariant products of collinear fields.  In some cases
perturbative matching computations are not necessary, since RPI gives relations
between Wilson coefficients that are valid to all orders in perturbation theory.

Heavy-to-light currents, $J= \bar q\Gamma b$, are important for describing a
broad range of processes with SCET, including both inclusive semileptonic and
radiative decays like $B\to X_u\ell\bar\nu$ and $B\to
X_s\gamma$~\cite{Bauer:2000ew,Bauer:2000yr,Bauer:2001yt,Bauer:2003pi,Bosch:2004th,Lee:2004ja,Bosch:2004cb,Beneke:2004in,Lange:2005yw,Chay:2005ck},
exclusive semileptonic and radiative decays such as $B\to \pi\ell\bar\nu$ and
$B\to
K^*\gamma$~\cite{Bauer:2000yr,Chay:2002vy,Beneke:2002ph,Bauer:2002aj,Pirjol:2002km,Chay:2003kb,Beneke:2003pa,Lange:2003pk,Grinstein:2004uu,Becher:2005fg},
and exclusive hadronic decays like
$B\to\pi\pi$~\cite{Chay:2003ju,Bauer:2004tj,Beneke:1999br,Feldmann:2004mg}.

Here we will consider higher order RPI relations for heavy-to-light currents in
a theory \SCETa with ultrasoft (usoft) and collinear fields.  Any momentum can
be decomposed as $p^\mu= n\mcdot p\: \bn^\mu/2+\bn\mcdot p\: n^\mu/2 +
p_\perp^\mu$, where $n_\mu, \bn_\mu$ are two light-cone vectors satisfying $n^2
= \bn^2 = 0$ and $n\cdot \bn = 2$.  The vector $n^\mu$ appears as a label for
the collinear quarks $\xi_n$ and gluons $A_n^\mu$, and the quantum fluctuations
described by these fields are predominately about this direction.  The collinear
modes have momentum scaling as $(n\cdot p, \bn\cdot p, p_\perp) \sim
Q(\lambda^2, 1, \lambda)$. The usoft modes $q_{us}, A_{us}^\mu$ have momenta
$p_{us}^\mu \sim Q\lambda^2$. We also use HQET usoft fields $h_v$ for heavy
quarks, where $v^\mu$ is a velocity label vector with $v^2=1$ (see for
example~\cite{Manohar:2000dt,Neubert:1993mb}).  The mass of the heavy quark is
denoted by $m$, $Q$ is a hard energy scale $\sim m$, and $\lambda\ll 1$ is the
SCET expansion parameter.  The auxiliary vectors $n,\nb$ and $v$ break part of
the full Lorentz symmetry of QCD, and this symmetry is restored order by order
in the power counting by reparameterization invariance under changes in these
parameters. For processes involving heavy-to-light currents it is often
convenient to work in the special frame where $v_\perp=0$, so that
$v^\mu=\bn\mcdot v\: n^\mu/2 + n\mcdot v\: \bn^\mu/2$ and $n\mcdot v\: \bn\mcdot
v=1$.

In HQET it is convenient to formulate the RPI constraints~\cite{Luke:1992cs} to
all orders in $1/m$ by constructing RPI invariant operators and then expanding
them to generate a chain of related operators. These operators start at some
fixed order in $1/m$, but once the RPI invariant form of this operator is known,
all higher terms in the chain are determined.  The RPI symmetries in SCET are
richer and typically the constraints are derived order by order in
$\lambda$. In this case, higher order operators in the chain are not fully
determined until the appropriate order in $\lambda$ is considered. 

Reparamaterization invariance constraints in SCET were first considered by Chay
and Kim~\cite{Chay:2002vy}.  The complete set of SCET RPI transformations were
formulated in Ref.~\cite{Manohar:2002fd} and used to prove that the leading
order (LO) SCET Lagrangian is not renormalized to all orders in perturbation
theory.  RPI constraints on subleading Lagrangians and tree level currents to
${\mathcal O}(\lambda^2)$ were derived in Ref.~\cite{Beneke:2002ph} (and
verified in~\cite{Lee:2004ja} for a basis with $v_\perp\ne 0$).  At ${\mathcal
  O}(\lambda)$, the extension to a complete set of heavy-to-light currents
constrained by RPI relations including currents that appear beyond tree-level
was made in Ref.~\cite{Pirjol:2002km}.  At this order, all Wilson coefficients
are constrained by RPI except for one scalar, four vector, and six tensor
currents, for which the one-loop matching was done in Ref.~\cite{Beneke:2004rc}
and independently in Ref.~\cite{Becher:2004kk}.  For the currents that survive
for $v_\perp=0$ the ${\mathcal O}(\lambda)$, RPI relations were verified in
Ref.~\cite{Hill:2004if}.  To simplify the computation, they considered
constraints restricted to the projected $v_\perp=0$ surface (from the
RPI-$\star$ transformation defined later) since this involves writing down 
fewer operators. At ${\mathcal O}(\lambda^2)$, the allowed set of field
structures for the heavy-to-light currents was determined in
Ref.~\cite{Beneke:2004in}. Four quark operator currents first appear at this
order.\footnote{In the most common decomposition the Wilson coefficients of the
  four quark operators start at ${\mathcal O}(\alpha_s^2)$, so these operators are
  not needed if the basis is restricted to LO in $\alpha_s(m_b)$, such as in
  Ref.~\cite{Lee:2004ja}.}  Recently the type-II RPI invariance was extended to
include light quark mass effects and provide constraints on certain $m_q$
dependent operators~\cite{Chay:2005ck}. For heavy-to-light currents at ${\cal
  O}(\lambda^2)$ a complete basis of Dirac structures and the full set of RPI
relations have not yet been constructed.

In constructing subleading operators we combine objects that are individually
collinear and usoft gauge invariant. The logic which ensures that all subleading
operators can be organized in terms of these objects relies on the decoupling of
usoft gluons from the leading order collinear Lagrangian by a field redefinition
involving a Wilson line $Y$~\cite{Bauer:2001yt}. In section~\ref{sect_Ybc} we
show that all results are independent of the choice of boundary condition for
this Wilson line. Processes described by SCET can depend on the path of Wilson
lines, but this path is determined independent of the choice of boundary
condition.

Our main objective in this paper is to to derive the complete basis of currents
at ${\mathcal O}(\lambda^2)$ by constructing a basis that is valid at any order
in perturbation theory and including all RPI relations.  Results are derived for
use in the $v_\perp=0$ frame (and we take $m_q=0$ in all currents).  Two
combinations of \{SCET RPI-I, SCET RPI-II, HQET RPI\} are formed which leave
$v_\perp=0$, and are called RPI-$\star$ and RPI-\$ (section~\ref{sect_rpi}). We
call these transformations on the surface $v_\perp=0$ ``projected RPI
relations'' and study their completeness in the full space of allowed
transformations in section~\ref{sect_complete}. For the ${\mathcal
  O}(\lambda^2)$ heavy-to-light currents, we show that transformations on the
projected surface give the complete set of relations for currents defined on
this surface (see section~\ref{sect_constD}). By eliminating the field operators
we show that it is convenient to consider the RPI relations as constraint
equations of the form
\begin{eqnarray} \label{constraint}
 \sum_{i,k} B_i(\omega_k)\ \Gamma^B_i  
    = \sum_{j,\ell} C_j(\omega_\ell)\ \Gamma^C_j \,,
\end{eqnarray}
where $B_i$ and $\Gamma^B_i$ are Wilson coefficients and Dirac structures for
operators that appear at some fixed order in $\lambda$, and $C_j$
and $\Gamma^C_i$ are terms that appeared in operators from lower orders.  By
deriving these constraint equations in section~\ref{sect_constB} prior to
searching for their solutions, it becomes easier to simultaneously consider the
restrictions imposed by the five different types of RPI invariance from both
SCET and HQET, since each gives a separate constraint. A simple counting
procedure is given for determining all possible Dirac structures prior to
imposing the RPI conditions. The solution of the constraint equations in
section~\ref{sect_constC} give relations between the $B_i$ and $C_j$
coefficients and determine the allowed Dirac structures $\Gamma^B_i$
in terms of $\Gamma^C_j$.

%%%%%%%%%%%%%%%%%%%%%%%%%%%%%%%%%%%%%%%%%%%%%%%%%%%%%%%%%%%%%%%%%%%%%%%%
\section{Ingredients from SCET}
%%%%%%%%%%%%%%%%%%%%%%%%%%%%%%%%%%%%%%%%%%%%%%%%%%%%%%%%%%%%%%%%%%%%%%%%

\subsection{Degrees of freedom, power counting, gauge invariance, and Wilson 
lines}

We briefly review some basic definitions from SCET that we will need for our
computations. The fields include collinear gluons $A_n^\mu$, ultrasoft gluons
$A_{us}^\mu$, collinear quarks $\xi_n$, and heavy quarks $h_v$. An important
attribute of our collinear fields is that they carry both a large label momentum
$p$ and a coordinate $x$, such as $\xi_{n,p}(x)$.  The label momenta are picked
out by momentum operators, $\bnP \xi_{n,p} = \bn\mcdot p\, \xi_{n,p}$ and ${\mathcal
  P}_\perp^\mu \xi_{n,p}= p_\perp^\mu \xi_{n,p}$ (see Ref.~\cite{Bauer:2001ct}),
while derivatives $i\partial^\mu$ act on $x$ and scale like ultrasoft momenta.
We define collinear covariant derivatives
\begin{eqnarray}
  i\nb\cdot D_c=\pb+g\nb\cdot A_{n}\,,\qquad
  iD_c^{\perp\mu} = \mathcal{P}_\bot^\mu +
  gA_{n}^{\perp\mu}\,,\qquad
 in\cdot D_c = i n\cdot \partial + g n\cdot A_n ,
\end{eqnarray}
and an usoft covariant derivative
\begin{eqnarray}
  iD_{us}^\mu = i\partial^\mu + gA_{us}^\mu.
\end{eqnarray}
To construct gauge invariant structures, it is useful to define
the collinear Wilson line
\begin{eqnarray} \label{W}
  W &\equiv& \Big[ \sum_{\textrm{perms}} \exp\Big( -\frac{g}{\pb}\: \nb\cdot A_n \
  \Big) \Big]
\end{eqnarray}
and an ultrasoft Wilson line 
\begin{eqnarray} \label{Y}
 Y(x^\mu) \equiv {\tilde {\textrm{P}}} 
  \exp \Big( i\, g \int^0_{s_0}
 \!\!
  ds \ n \mcdot
 A_{us}(x^\mu + s n^\mu ) \Big) \,,
\end{eqnarray}
where it is convenient to choose the reference point $s_0$ to be $s_0=-\infty$
with $\tilde {\textrm P}={\textrm P}$ path ordering for both quarks and
antiquarks. We comment on the $s_0$ independence of results in the next
section.  Making the collinear field redefinitions~\cite{Bauer:2001yt}
\begin{eqnarray} \label{fd}
 \xi_{n,p}(x) \to Y(x) \xi_{n,p}(x) \,,\qquad 
 A_{n,q}(x) \to Y(x) A_{n,q}(x) Y^\dagger(x) \,,
\end{eqnarray}
removes all couplings to usoft gluons from the leading order collinear
Lagrangian and induces factors of $Y$ in operators (giving a simple statement of
usoft-collinear decoupling).  

We will use the following structures, which are both collinear and usoft gauge
invariant:
\begin{equation} \label{objs}
\chi_n \equiv W^\dagger \xi_n , \qquad 
 {\mathcal H}_v \equiv Y^\dagger h_v, \qquad
{\mathcal D}_c \equiv W^\dagger D_c W \,,  
\quad\quad {\mathcal D}_{us} \equiv Y^\dagger D_{us} Y \,,
\end{equation}
as well as the $\cP_\perp^\mu$ label momentum operator. The fields in
Eq.~(\ref{objs}) are all post-field redefinition.  It is convenient to be able
to switch the collinear derivatives for field strengths, for which we use
\begin{eqnarray} \label{toBD}
   i{\mathcal D}_c^{\!\perp\mu} 
  &=& \cP_\perp^\mu  + ig \cBp^\mu \,, \qquad\qquad\!\!\!
  i\overleftarrow {\mathcal D}_c^{\!\perp\mu} 
  = - \cP_\perp^{\,\dagger\mu} - ig \cBp^\mu   \,, \nn\\[6pt]
  i n\mcdot {\mathcal D}_c &=& in\mcdot\partial + i g n\mcdot \cB \,,\qquad
  i n\mcdot \overleftarrow{\mathcal D}_c 
   =in\mcdot\overleftarrow\partial - i g n\mcdot \cB \,.
\end{eqnarray}
Here the field strength tensors are
\begin{equation}
 i g \cBp^\mu \equiv \Big[\frac{1}{\pb} [i\nb\cdot {\mathcal D}_c,i
{\mathcal D}_c^{\perp\mu}] \Big] \,, \qquad\qquad 
 i g n \cdot \cB \equiv \Big[\frac{1}{\pb}  [i\nb\cdot {\mathcal D}_c,i n
\cdot {\mathcal D}_c] \Big],
\end{equation}
where the label operators and derivatives act only on fields inside the outer
square brackets. To determine which fields appear in the heavy-to-light current
at each order of $\lambda$, we need the $\lambda$-scaling of the operators
listed in Table \ref{table_pc}.

For convenience we will also use the shorthand notation
\begin{eqnarray} \label{sh}
 \bar \chi_{n,\omega}
  &\equiv& \Big[\bar \chi_n \: \delta(\omega\!-\!n\cdot v\pb^\dagger) \Big]\,, 
   \nn\\
 ( i g \cBp^\mu)_{\omega}
 &\equiv& \Big[ i g \cBp^\mu\:
  \delta(\omega\!-\!n\cdot v \pb^\dagger) \Big]\,, \nn \\
 ( i g n \cdot \cB)_{\omega}
 &\equiv& \Big[ i g n \cdot \cB\:
  \delta(\omega\!-\!n\cdot v \pb^\dagger) \Big]\,,
\end{eqnarray}
so that $\bar \chi_{n,\omega}$ corresponds to the gauge invariant combination of
fields $(\bar\xi_n W)$ carrying large ${\mathcal O}(\lambda^0)$ momentum
$\omega$. An operator built out of several of these components then has multiple
labels, $J(\omega_1,\omega_2,\ldots)$, and the Wilson coefficient for the
operator will be a function of the same $\omega_i$ momentum labels,
$C(\omega_1,\omega_2,\ldots)$.

\begin{table}[t!]
\begin{center}
\begin{tabular}{l|c|cc|cc|cccc}
 &  collinear quark & soft & quarks & label & operators & & covariant &  derivatives &
  \\\hline Operator &  $\xi_n $ & $h_v$& $q_{us}$ & $\pb$ & $\mathcal{P}_\bot^\mu$ & $i \nb \cdot D_c$& $iD_c^{\perp \mu}$ & $i n \cdot D_c$ & $iD_{us}^\mu$ \\ Scaling &
   $\lambda$ &$\lambda^3$&$\lambda^3$ & $\lambda^0$ & $\lambda$ & $\lambda^0$ & $\lambda$ & $\lambda^2$ & $\lambda^2$
\end{tabular}
\end{center}
{\caption{Power counting for effective theory operators.}
\label{table_pc} }
\end{table}
We will use a $T$ subscript or superscript to denote objects transverse to
$v^\mu$, and a $\perp$ to denote those perpendicular to $n^\mu$ and $\bn^\mu$,
\begin{eqnarray}
  R^\mu_T = R^\mu - v^\mu v\mcdot R \,,\qquad\qquad
  R^\mu_\perp =R^\mu-\frac{n^\mu}{2}\: \bn\mcdot R -\frac{\bn^\mu}{2}\: n\mcdot R
  \,.
\end{eqnarray}
The effective theory fields satisfy the projection relations $P_n \chi_n =
\chi_n$, $\cnW_n P_{\nb}=\cnW_n$, and $P_v {\mathcal H}_v={\mathcal H}_v$ where the
matrices are
\begin{eqnarray}
  P_n =\frac{\slashed{n}\nbs}{4}\, , 
  \quad P_{\nb}=\frac{\nbs \slashed{n}}{4}\, , \quad \text{and} \quad
  P_v=\half\bigl(1+\slashed{v} \bigr) \,.
\end{eqnarray} 
The number of independent Dirac structures in a current is reduced by these
relations. For $\cnW_n \Gamma {\mathcal H}_v$, we can project the Dirac structure
onto a four dimensional basis $\{1, \gamma^5, \gammapa\}$ using
\begin{eqnarray}  \label{reduce1}
  \Gamma
  \doteq \tr{\left[ P_{\nb} \Gamma P_v \right]} +\gamma^5 \tr{\left[ \gamma^5
      P_{\nb} \Gamma P_v \right]} +\gamma^\bot_\alpha \tr{\left[ \gammapa
      P_{\nb} \Gamma P_v \right]} \,,
\end{eqnarray}
where $\doteq$ indicates that the relation is true between $\cnW_n$ and ${\mathcal
  H}_v$.  Similarly, between collinear quark fields, $\cnW_n \Gamma \chi_n$,
we can project the Dirac structure onto the basis $\{\nbs, \nbs \gamma^5, \nbs
\gammapa\}$ using
\begin{eqnarray} \label{reduce2}
    \Gamma \doteq \frac{\nbs}{8} \tr{\left[ \ns P_{\nb} \Gamma P_n \right]}
    +\frac{\nbs\gamma^5}{8} \tr{\left[ \gamma^5 \ns P_{\nb} \Gamma P_n \right]}
    +\frac{\nbs\gamma^\bot_\alpha}{8} \tr{\left[ \gammapa \ns P_{\nb}
    \Gamma P_n \right]} \,.
\end{eqnarray}
Finally we define $\epsilon_\perp^{\mu\nu}=\nb_\rho n_\sigma
\epsilon^{\mu\nu\rho\sigma}/2$ where $\epsilon_\perp^{12}=+1$, and note that the
tensor identity,
\begin{equation}
    g_\bot^{\alpha[\mu}\epsilon_\bot^{\nu]\beta} =
    -g_\bot^{\alpha\beta}\epsilon_\bot^{\mu\nu} \,,
\end{equation}
will be useful.

%\newpage
%%%%%%%%%%%%%%%%%%%%%%%%%%%%%%%%%%%%%%%%%%%%%%%%%%%%%%%%%%%%%%%%%%%%%%
\subsection{Comments on boundary conditions for $Y(x)$} \label{sect_Ybc}

It is worth making a few comments on the path and $s_0$ dependence of the
ultrasoft Wilson lines used in Eq.~(\ref{fd}).  This field redefinition is
universal and should apply equally well for any physical process. We define
\begin{eqnarray} \label{Yd}
   Y(x^\mu) &=& { \tilde {\textrm{P}}} 
  \exp \Big( i g \int^0_{s_0} \!\!
 % _{-\infty\, {\textrm{sign}}(\bnP) }0 \!\!\!\!\!\!\!\!\!\!\!\!\!\!\!
  ds \ n \mcdot
 A_{us}(x_s^\mu) \Big)\,, \\[5pt]
Y^\dagger(x^\mu) &=& { {\tilde {\textrm{P}}'} } 
    \exp \Big( \!-\! i g \int^0_{\overline s_0}
 %_{\infty\, {\textrm{sign}}(\bnP^\dagger ) }
 % \!\!\!\!\!\!\!\!\!\!\!\!\!\!\!
  ds \ n \mcdot
 A_{us}(x_s^\mu) \Big) \,, \nn
\end{eqnarray}
where $x_s^\mu = x^\mu + s n^\mu $.  With respect to the equation of motion,
$n\mcdot D_{us} Y=0$, the point $s_0$ implements a boundary condition at
infinity, and $\tilde {\textrm P}$ denotes path ordering ${\textrm P}$ or
anti-path ordering $\overline {\textrm P}$.  If $Y^\dagger$ is to be the
hermitian conjugate of $Y$ one requires that $\overline s_0=s_0^\dagger$ and
$\tilde {\textrm P}' = \overline {\tilde {\textrm P}}$.  This ensures that
$Y^\dagger Y = 1$ and that the field redefinition in Eq.~(\ref{fd}) causes the
usoft gluons to decouple in the collinear Lagrangian.  The following definitions
will also be useful
\begin{align}
  Y_+  &= {\textrm{P}} 
    \exp \Big( ig\! \int^0_{-\infty }
 \!\!\!\!\!
 %\!\!\!\!\!\!\!\!\!\!\!\!\!\!\!
  ds \ n \mcdot
 A_{us}(x_s^\mu ) \Big) 
 \,,
  & Y_- &= \overline {\textrm{P}} 
    \exp \Big( \!\!-\!\! ig\! \int_0^{\infty }
 \!\!\!\!\!
 %\!\!\!\!\!\!\!\!\!\!\!\!\!\!\!
  ds \: n \mcdot
 A_{us}(x_s^\mu) \Big)
   \,, \\
  Y_-^\dagger  &=  \overline {\textrm{P}} 
    \exp \Big(\!\!-\!\! ig\! \int^0_{-\infty }
 \!\!\!\!\!\!
 %\!\!\!\!\!\!\!\!\!\!\!\!\!\!\!
  ds \: n \mcdot
 A_{us}(x_s^\mu  ) \Big)  
  \,, 
 & Y^\dagger_+  
 &= {\textrm{P}} 
    \exp \Big(  ig\! \int_0^{\infty }
 \!\!\!\!\!
 %\!\!\!\!\!\!\!\!\!\!\!\!\!\!\!
  ds \ n \mcdot
 A_{us}(x_s^\mu  ) \Big)
   \,. \nn
\end{align}
Here $(Y_\pm)^\dagger = Y_\mp^\dagger$, and the subscript on $Y_\pm^\dagger$
should be read as $(Y^\dagger)_\pm$ rather than $(Y_\pm)^\dagger$.

A common choice for $s_0$ is the one made in Ref.~\cite{Bauer:2001yt},
\begin{eqnarray} \label{choice1}
  s_0= \overline s_0 = -\infty \,, \qquad
  \tilde {\mathrm P}={\mathrm P} \,,\quad
  {\tilde{\mathrm P}}^\prime=\overline {\mathrm P}\,,
\end{eqnarray}
where $Y=Y_+$ and $Y^\dagger=Y_-^\dagger$. In Ref.~\cite{Bauer:2003di} the
choice $s_0=+\infty$ with $\tilde {\textrm P}'= {\textrm P}$ was made in
order to correspond with particle production, $Y^\dagger=Y_+^\dagger$. A third
possible choice is~\cite{Bauer:2002aj}
\begin{eqnarray} \label{thechoice}
 s_0= -\infty \,{\textrm{sign}}(\bnP) \,,\quad
 \overline s_0 = -\infty \,{\textrm{sign}}(\bnP^\dagger)\,, \quad
  \left\{ \begin{matrix}
    \tilde {\textrm P} \!=\! {\textrm P} \,,
      \tilde {\textrm P}' \!=\!  \overline {\textrm P}  
      \ \ {\textrm {for}}\ \  \bnP , \bnP^\dagger> 0 \\
    \tilde {\textrm P} \!=\! \overline {\textrm P} \,,
      \tilde {\textrm P}' \!=\! {\textrm P} 
      \ \ {\textrm {for}}\ \  \bnP , \bnP^\dagger < 0 
   \end{matrix}  
   \right. \,.
\end{eqnarray}
Eq.~(\ref{thechoice}) still satisfies $s_0^\dagger = \overline s_0$ but
corresponds to a different choice for particles and antiparticles.\footnote{Note
  that in this case $s_0=-\infty\: {\mathrm{ sign}}(\bnP)$, is an operator.
  } Here $Y=Y_+$, $Y^\dagger =Y_-^\dagger$ for particles, while $Y=Y_-$,
$Y^\dagger=Y_+^\dagger$ for antiparticles.  To see this recall that
\begin{eqnarray}
  \xi_{n,p} =  \xi_{n,p}^+  +  \xi_{n,-p}^- \,,
\end{eqnarray}
and that if the label momentum is positive $\bn\mcdot p>0$ we get the field for
particles, $\xi_n^+$, and if the label is negative $\bn\mcdot p<0$ we get the
field operator for antiparticles, $\xi_n^-$~\cite{Bauer:2001ct}.  Although it is
important to make some choice for $s_0$, if one is careful then in any physical
problem the dependence on $s_0$ cancels. Any path dependence exhibited by a
final result can be derived independently of the choice of $s_0$ that one makes in the field redefinition.

Since the dependence on $s_0$ sometimes causes confusion, we explore some of the
subtleties in this section, in particular, why it is important to remember that
factors of $Y$, $Y^\dagger$ can also be induced in the interpolating fields for
incoming and outgoing collinear states, and why a common choice for $s_0 =
\overline s_0^{\,\dagger}$ is sufficient to properly reproduce the $i\epsilon$
prescription in perturbative computations. In many processes (examples being
color allowed $B\to D\pi$ and $B\to X_s\gamma$) the $s_0$ dependence of the
Wilson lines cancels and the following considerations are not crucial. In other
processes, however, the path for the Wilson line is important for the final
result, particularly when these Wilson lines do not entirely cancel. An example
of this is jet event shapes as discussed in
Refs.~\cite{Korchemsky:1999kt,Bauer:2002ie,Bauer:2003di}.  See also the
discussion of path dependence in eikonal lines in
Refs.~\cite{Collins:1981ta,Bodwin:1984hc,Collins:1988ig,Korchemsky:1992xv,Collins:2002kn,Ji:2004wu,Collins:2004nx,Chay:2004zn}.

\begin{figure} 
%\begin{center}
%
\includegraphics[height=1.7cm]{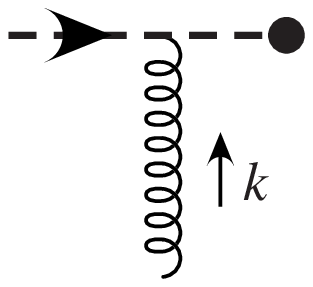} 
  \raisebox{20pt}{\LARGE $\frac{i}{n\cdot k + i\epsilon}$}\ \
\includegraphics[height=1.7cm]{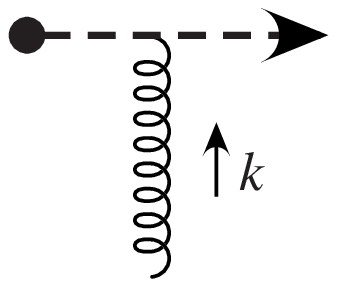} \!\!
  \raisebox{20pt}{\LARGE $\frac{i}{-n\cdot k + i\epsilon}$}\ \ 
\includegraphics[height=1.7cm]{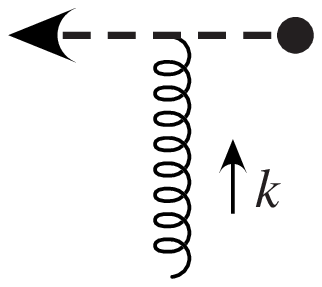} 
  \raisebox{20pt}{\LARGE $\frac{i}{-n\cdot k - i\epsilon}$}\ \ 
\includegraphics[height=1.7cm]{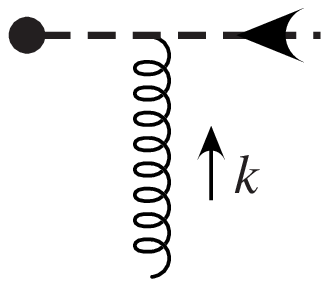} 
  \raisebox{20pt}{\LARGE $\frac{i}{n\cdot k - i\epsilon}$} \\
\vskip2pt
\hspace{-0.5cm}
$(Y_+\, \xi_n^+)$\hspace{2.5cm} $(\bar\xi_n^+\, Y_+^\dagger)$\hspace{3.cm}
$(\bar \xi_n^-\, Y_-^\dagger)$ \hspace{2.7cm} $(Y_-\, \xi_n^-)$
%\vskip5pt
%\end{center}
\vskip0pt
\caption[1]{
  Eikonal $i\epsilon$ prescriptions for incoming/outgoing quarks and antiquarks
  and the result that reproduces this with an ultrasoft Wilson line and sterile
  quark field.
\label{fig_Y}}
\end{figure}
First consider the perturbative computation of attachments of usoft gluons to
incoming and outgoing quark and antiquark lines. The results for the eikonal
factors for one gluon are summarized in Fig.~\ref{fig_Y}, and can be computed
directly with the SCET collinear quark Lagrangian (or from an appropriate limit
of the QCD propagator). These attachments seem to force one to make a particular
choice for $s_0$ and $\overline s_0$, see for example the recent detailed study
in Ref.~\cite{Chay:2004zn}. In our notation it is straightforward to show that
this choice corresponds to
\begin{eqnarray} \label{theCchoice}
 s_0=-\infty\, \textrm{sign}(\bnP) \,,\quad 
 \overline s_0 = +\infty\, \textrm{sign}(\bnP^\dagger) \,,\quad
 \left\{ \begin{matrix}
    \tilde {\textrm P} \!=\! \tilde {\textrm P}' \!=\! {\textrm P} \,,
      \ \ {\textrm {for}}\ \  \bnP , \bnP^\dagger> 0 \\
    \tilde {\textrm P} \!=\! \tilde {\textrm P}' \!=\! \overline {\textrm P} \,,
      \ \ {\textrm {for}}\ \  \bnP , \bnP^\dagger < 0 
   \end{matrix} 
  \right. \,.
\end{eqnarray}
To see this take a quark with label $\bn\mcdot p>0$ and an antiquark with label
$\bn\mcdot p' <0$, and note that
\begin{align} \label{Yc1}
  Y \xi_{n,p} &= \tilde {\textrm{P}} 
    \exp \Big( ig\! \int^0_{-\infty }
 \!\!\!\!\!
 %\!\!\!\!\!\!\!\!\!\!\!\!\!\!\!
  ds \ n \mcdot
 A_{us}(x_s^\mu)  \Big) \xi_{n,p}^+  = {\textrm{P}} 
    \exp \Big( ig\! \int^0_{-\infty }
 \!\!\!\!\!
 %\!\!\!\!\!\!\!\!\!\!\!\!\!\!\!
  ds \ n \mcdot
 A_{us}(x_s^\mu ) \Big) \xi_{n,p}^+ 
 \equiv Y_+ \xi_{n,p}^+ \,,  \\
\bar\xi_{n,p} Y^\dagger \! &= \bar\xi_{n,p}^+ \tilde {\textrm{P}}'
    \exp \Big( \!-\! ig\! \int^0_{\infty }
 \!\!\!\!\!
 %\!\!\!\!\!\!\!\!\!\!\!\!\!\!\!
  ds \ n \mcdot
 A_{us}(x_s^\mu ) \Big) \! 
 = \bar\xi_{n,p}^+ {\textrm{P}} 
    \exp \Big(  ig\! \int_0^{\infty }
 \!\!\!\!\!
 %\!\!\!\!\!\!\!\!\!\!\!\!\!\!\!
  ds \ n \mcdot
 A_{us}(x_s^\mu  ) \Big)
 \equiv \bar\xi_{n,p}^+ Y^\dagger_+    \,,\nn\\
Y \xi_{n,p'}\! &= \tilde {\textrm{P}} 
    \exp \Big( ig\!\! \int^0_{\infty }
 \!\!\!\!
 %\!\!\!\!\!\!\!\!\!\!\!\!\!\!\!
  ds \: n \mcdot
 A_{us}(x_s^\mu ) \Big) \xi_{n,p'}^- 
  = \overline {\textrm{P}} 
    \exp \Big( \!\!-\!\! ig\! \int_0^{\infty }
 \!\!\!\!\!
 %\!\!\!\!\!\!\!\!\!\!\!\!\!\!\!
  ds \: n \mcdot
 A_{us}(x_s^\mu) \Big) \xi_{n,p'}^- 
  \equiv  Y_- \xi_{n,p'}^- \,, \nn \\
\bar\xi_{n,p'} Y^\dagger \! &= \bar\xi_{n,p'}^- \tilde {\textrm{P}}'
    \exp \Big( \!\!-\!\! ig\! \int^0_{-\infty }
 \!\!\!\!\!\!
 %\!\!\!\!\!\!\!\!\!\!\!\!\!\!\!
  ds \: n \mcdot
 A_{us}(x_s^\mu  ) \Big) \! 
 = \bar\xi_{n,p'}^- \overline {\textrm{P}} 
    \exp \Big(\!\!-\!\! ig\! \int^0_{-\infty }
 \!\!\!\!\!\!
 %\!\!\!\!\!\!\!\!\!\!\!\!\!\!\!
  ds \: n \mcdot
 A_{us}(x_s^\mu  ) \Big)  
  \equiv \bar \xi_{n,p'}^- Y_-^\dagger \,.\nn
\end{align}
This is in agreement with the $\tilde Y = Y_-$, $Y^\dagger= Y_-^\dagger$,
$Y=Y_+$, $\tilde Y^\dagger = Y_+^\dagger$ used in~\cite{Chay:2004zn} for the
production and annihilation of antiparticles and the annihilation and production
of particles respectively.  The results in Eq.~(\ref{Yc1}) reproduce the natural
choice of having incoming quarks/antiquarks enter from $-\infty$, while outgoing
quarks/antiquarks extend out to $+\infty$. 

Although the choice in Eq.~(\ref{theCchoice}) agrees with the $i\epsilon$'s in
Fig.~\ref{fig_Y} it causes complications in the attachments of usoft gluons to
internal collinear propagators. With Eq.~(\ref{theCchoice}) we have $s_0^\dagger
\ne \overline s_0$. Now the field redefinition still induces factors of
$Y^\dagger_+ Y_- = 1$ and $Y^\dagger_- Y_+ = 1$ in production and annhilation
terms in the collinear Lagrangian, but it also induces factors
of $Y^\dagger_+ Y_+ = Y_\infty $ and $Y_-^\dagger Y_- = Y_\infty^\dagger$ in
quark-quark and antiquark-antiquark terms in the action, where
\begin{eqnarray}
  Y_\infty = {\textrm{P}} 
    \exp \Big( ig\! \int^{+\infty}_{-\infty }
 \!\!\!\!\!
  ds \ n \mcdot
 A_{us}(x_s^\mu)  \Big)\,.
\end{eqnarray}
When usoft gluons attach to a collinear propagator with endpoints $x$ and $y$ we
must end up with a finite Wilson line $Y(x,y)$. In the original collinear
Lagrangian (prior to the field redefinition) this finite Wilson line is
generated by the time ordering of fields in the usoft gluon interaction
vertices. If a field redefinition is made with boundary conditions satifying
$s_0^\dagger = \overline s_0$ then the vertices bordering a collinear
propagator induce Wilson lines whose $s_0$ dependence cancels, leaving this same
finite Wilson line. For example, with $s_0=-\infty$,
$Y(-\infty,x)Y(-\infty,0)^\dagger=Y(0,x)$.  A choice like that in
Eq.~(\ref{theCchoice}) is more complicated since it violates hermiticity:
$(\xi_n)^\dagger =\xi_n^\dagger $ prior to the field redefinition, but this is
no longer true for the $\xi_n$ and $\bar\xi_n$ fields after the field
redefinition. Correspondingly, the term in the action determining the free
propagator depends on $Y_\infty$.  Thus, in this case there are $Y$ factors in
both the propagators and vertices which must be taken into account in order for
the path ordering not to conflict with the result from time ordering, and give
the same finite Wilson line.

Let's adopt the choice in Eq.~(\ref{choice1}) rather than Eq.~(\ref{theCchoice})
and check that the theory with the field redefinition in Eq.~(\ref{fd}) still
correctly reproduces the results in Fig.~\ref{fig_Y} for this case. Here we have
$Y=Y_+$, $Y^\dagger = Y_-^\dagger$ for particles and antiparticles. Thus, the
correct $i\epsilon$'s are obviously reproduced for the incoming collinear lines
as well as intermediate propagator states. On the other hand, the result for an
outgoing quark seems to have the wrong factor since $\bar\xi_n^+$ comes with a
$Y_-^\dagger$ rather than a $Y_+^\dagger$. However, with the standard definition
of an outgoing state there is actually an extra $Y_\infty$ induced by the field
redefinition on the out-state itself. When we take this factor into account we
have $Y_\infty Y_-^\dagger = Y_+^\dagger$ as expected. To see this, recall that
an outgoing collinear quark state ${}_{\textrm out}\langle \vec p |$ is
generated by a suitably weighted integral over $\langle 0 | \xi_n^+(x_T)$, in
the large time limit $T\to\infty$ for $x_T=(T,\vec x)$. When we make the field
redefinition this field, $\xi_n^+(x_T)$ generates an usoft Wilson line which
extends from our reference point $s_0=-\infty$ to the $\bn\mcdot x$ point for
our asymptotic outgoing state (which is $+\infty$ for $T\to\infty$), namely a
factor of $Y_\infty$. A similar argument applies for outgoing antiquark states,
where we get $Y_+ Y_\infty^\dagger=Y_-$.  The same considerations must also be
made for hadronic bound states where they apply to the interpolating
quark/antiquark fields used along with the LSZ formula to define the outgoing
state. The factors of $Y_\infty$ are universal, independent of which out-state
we choose.  There are no additional factors for our incoming states since our
reference point and $T=-\infty$ coincide, $Y(-\infty,-\infty)=1$. Once the
$Y_\infty$ factors are taken into account, the choice in Eq.~(\ref{choice1})
correctly reproduces the path for outgoing quark and antiquark lines. If we had
instead made the choice for $s_0$ in Eq.~(\ref{thechoice}) (which also satisfies
$s_0^\dagger = \overline s_0$) then we would have $Y_\infty^{(\dagger)}$ factors
for incoming antiquark states and outgoing quark states, but the final outcome
is the same. Thus the complete result is independent of the $s_0$ choice.

The above discussion covers usoft interactions from the collinear Lagrangian,
but it is also worth remarking on the interactions induced by the field
redefinition in (possibly non-local) operators that are not time ordered. We
continue to use Eq.~(\ref{choice1}). Here again, the identity $Y^\dagger Y = 1$
is important in order to prove the cancellation of usoft gluon attachments.  It
is convenient to adopt a convention where one collects the extra factors of
$Y_\infty$ induced from outgoing states together with the $Y^\dagger$'s from
production fields in these operators.  In this case if we consider $J(x)=\bar
\xi_{n}^+ \,\bnslash\, \xi_n^-$ for the production of a collinear quark and
antiquark, then instead of writing only the $Y_-^\dagger$ and $Y_+$ from the
fields we write $J\to \bar\xi_n^+ Y_+^\dagger \,\bnslash\, Y_- \,\xi_n^-$ =
$\bar\xi_n^+ \,\bnslash\,\xi_n^-$ which includes the $Y$'s from any out-state
this current could produce. Here the usoft interactions in the $Y$ and
$Y^\dagger$ lines extend from $x$ to $\infty$ and cancel.  For the annihilation
of a quark and antiquark the lines extend from $-\infty$ to $x$ and also cancel,
namely $Y^\dagger_- Y_+ =1$.  These two cancellations are often sufficient to
ensure the decoupling of usoft gluons.  For example, in exclusive processes we
must have color singlet combinations to connect to incoming or outgoing
collinear hadrons and so we can typically pair up $\bar\xi_n^\pm$ and
$\xi_n^\mp$ fields in the hard scattering operator and make the cancellations
manifest.

If we instead consider an inclusive process like DIS then we have a quark
scattered to a quark (we consider generic Bjorken $x<1$ in the Breit frame). In
this case including the $Y_\infty$ from one outgoing quark in the final state
gives $\bar\xi_n^+ \, \bnslash \, \xi_n^+ \to \bar\xi_n^+ Y_+^\dagger \,\bnslash
\, Y_+ \xi_n^+ $ where the Wilson lines do not seem to cancel. Here in order for
the cancellation of usoft gluons to take place it is important to either a) take
into account all factors of $Y_\infty$ from the outgoing proton state, or b)
include the $Y_\infty$ from one outgoing quark state but note that we are only
matching cut diagrams for this inclusive process. The choice a) or b) depends on
whether we want to take the imaginary part at the very end, or from the
beginning.  For b) the effective theory computation has the imaginary part of
the hard computation, but the imaginary part also effects the collinear
operator, where we can denote the cut by a vertical line, $\big|$.  With our
initial state for the $T$-matrix taken on the RHS of the cut, the signs are as
in Fig.~\ref{fig_Y}, but on the LHS we have the complex conjugate of these
expressions, and the above computation becomes
\begin{eqnarray}
 (\bar\xi_n) \big| ( \bnslash \xi_n)
    \to (\bar\xi_n Y_-^\dagger) \big| (\bnslash Y_+\xi_n) 
 % = (\bar\xi_n Y_+^\dagger Y_-) \big | (\bnslash \xi_n) 
  = \bar\xi_n \,\bnslash\, \xi_n \,.
\end{eqnarray}
Thus, the usoft gluon interactions also cancel in this case. Alternatively, with
a) one must keep track of all the lines in the full forward scattering
calculation including $Y_\infty^{(\dagger)}$ factors from all initial and/or
final state quarks, and then the $Y$'s in the low energy theory again all
cancel.  Both ways we arrive at the same final result, $({\textrm {Im}}\, C)
\bar\xi_n \bnslash \xi_n$ (see Refs.~\cite{Bauer:2002nz,Manohar:2003vb} for a
discussion of DIS in SCET).  Similar considerations can be applied to $B\to
X_s\gamma$ in the endpoint region. The $s_0$ dependence cancels, and for this
process we are left with a finite usoft Wilson line, $\bar h_v(x) Y(x,0)
h_v(0)$.
 
To summarize, keeping careful track of the boundary condition $s_0$ dependence
in the usoft Wilson line $Y$, a choice satisfying $s_0=\overline s_0^\dagger$
appears to be the most natural (even though there will be additional $Y_\infty$
factors from states). Physical results are independent of the choice made for
the $s_0$ reference point. They may still depend on the path of Wilson lines in
the final result, but this is determined by the universal class of processes
described by the operator rather than the choice of $s_0$ in the field
redefinition. Similar conclusions hold for the path dependence in collinear
Wilson lines $W$.  We note that with respect to the definitions of the gauge
invariant structures made in Eq.(\ref{sh}), the remaining allowed global color
rotations simply correspond to color rotations at the reference point. We will
pick the same reference point in $W$ and $Y$ factors. For example, the gauge
invariant product of fields $(Y^\dagger h_v)$ carries a color index in the {\bf
  3} representation, which by convention is acted on by global rotations
$U(s_0)$, via $(Y^\dagger h_v)\to U(s_0)(Y^\dagger h_v)$. These color rotations
still connect invariant products of collinear and usoft fields.

\subsection{Reparameterization invariance} \label{sect_rpi}

The structure of the currents is constrained by reparameterization invariance,
which is an invariance that appears due to the ambiguity in the decomposition of
momenta in terms of basis vectors and in terms of large and small components. 
The total momentum $P^\mu$ of a heavy quark is decomposed as
$$P^\mu = m_Q v^\mu + k^\mu,$$
where $m_Q$ is the quark's mass, $v^\mu$ is its
velocity, and $k^\mu$ is a residual momentum of order $m_Q \lambda^2$.  Then the
simultaneous shifts
\begin{equation}
v^\mu \rightarrow v^\mu + \beta^\mu \quad\text{and}\quad k^\mu
\rightarrow k^\mu - m_Q \beta^\mu,
\end{equation}
where the infinitesimal $\beta^\mu\sim \lambda^2$, can have no physical
consequences~\cite{Luke:1992cs}.  We refer below to this reparameterization
invariance as HQET-RPI.  The transformation of the field ${\mathcal H}_v\to {\mathcal
  H}_v + \delta {\mathcal H}_v$ induces terms at ${\mathcal O}(\lambda^0)$ and ${\mathcal
  O}(\lambda^2)$,
\begin{equation}
 \delta^{(\lambda^0)} {\mathcal H}_v = (i m \beta \cdot x ) {\mathcal H}_v 
   \,,\qquad\quad
 \delta^{(\lambda^2)} {\mathcal H}_v = 
    \frac{\beta\!\!\!\slash}{2}\: {\mathcal H}_v.
\end{equation}

There are also reparameterization invariances associated with ambiguities in the
decomposition of the momenta of collinear fields. Here the total momentum
$P^\mu$ of a collinear particle is decomposed into the sum of a collinear
momentum $p^\mu$, with $( n \cdot p, \nb \cdot p, p_\bot) \sim
Q(\lambda^2,1,\lambda)$, and an ultrasoft momentum $k^\mu$, with $( n \cdot k,
\nb \cdot k, k_\bot) \sim Q(\lambda^2,\lambda^2,\lambda^2)$:
\begin{eqnarray}
P^\mu &=& p^\mu + k^\mu \\
&=& \frac{n^\mu}{2} \nb \cdot (p + k) + \frac{\nb^\mu}{2} n \cdot
k + (p_\bot + k_\bot).
\end{eqnarray}
This decomposition has two types of ambiguity.  The first comes from splitting
$P^\mu$ into large ($p$) and small ($k$) components.  Thus operators must be 
invariant under a transformation that takes
\begin{eqnarray} \label{trnsfm-a}
&& \pb \rightarrow \pb + \nb \cdot \ell \,,\qquad\quad 
 i \nb\cdot \partial \rightarrow i \nb \cdot \partial - \nb \cdot \ell 
 \,,\nn\\
&& {\mathcal P}_\perp^\mu \rightarrow {\mathcal P}_\perp^\mu + \ell_\perp^\mu
\,, \qquad\quad 
 i\partial_\perp^\mu \rightarrow i\partial_\perp^\mu - \ell_\perp^\mu,
\end{eqnarray}
where all operators and derivatives act on one or more collinear fields, and
$\ell^\mu$ is ${\mathcal O}(\lambda^2)$. We refer to this reparameterization invariance as SCET
RPI-a. Examples of an infinitesimal transformation
on fields and operators are
\begin{eqnarray}
  \delta_a^{(\lambda^0)} \chi_n = (i \ell\cdot x) \chi_n \,,\qquad
  \delta_a^{(\lambda^1)} {\mathcal P}_\perp^\alpha = \ell^\alpha_\perp\,,\qquad
  \delta_a^{(\lambda^2)} \bnP = \bn\mcdot\ell \,,
\end{eqnarray}
where $n\mcdot\ell=0$.  Note that $(i\ell\cdot x)$ terms only effect ultrasoft derivatives
acting on the fields since the overall current is evaluated at $x=0$.

The second ambiguity in the decomposition of the momentum of the collinear
particles comes from choosing the light-cone vectors $n$ and $\nb$.  An
infinitesimal change in these vectors which preserves the relations $n^2 = 0$,
$\nb^2 = 0$, and $n \cdot \nb = 2$, can have no physical consequences.  The most
general infinitesimal transformations of $n$ and $\nb$ that preserve these
conditions along with the collinear power counting are~\cite{Manohar:2002fd}
\begin{eqnarray}\label{repinv}
\text{(I)} \left\{
\begin{tabular}{l}
$n_\mu \to n_\mu + \Delta_\mu^\perp$ \\
$\nb_\mu \to \nb_\mu$
\end{tabular}
\right.\qquad \text{(II)} \left\{
\begin{tabular}{l}
$n_\mu \to n_\mu$ \\
$\nb_\mu \to \nb_\mu + \varepsilon_\mu^\perp$
\end{tabular}
\right.\qquad \text{(III)} \left\{
\begin{tabular}{l}
$n_\mu \to (1+\alpha)\, n_\mu$ \\
$\nb_\mu \to (1-\alpha)\, \nb_\mu$
\end{tabular}
\right. \,,
\end{eqnarray}
where $\{\Delta^\bot_\mu, \epsilon^\bot_\mu, \alpha\}\sim
\{\lambda^1,\lambda^0,\lambda^0 \}$ are five infinitesimal parameters, and $\nb
\cdot \epsilon^\bot = n \cdot \epsilon^\bot = \nb \cdot \Delta^\bot = n \cdot
\Delta^\bot = 0$.  

If we start in the frame $v_\bot = 0$, then transformations (I) or (II) or
(HQET-RPI) take us out of this frame. A certain combined type I and type II
transformation, however, leaves $v_\perp=0$~\cite{Hill:2004if}. We refer to this transformation as RPI-$\star$.  We can also form a combined HQET and type II transformation
that leaves $v_\perp=0$ which we refer to as RPI-$\$$. These transformations are
\begin{eqnarray}\label{repinvnew}
(\star) \left\{
\begin{tabular}{l}
$n_\mu \to n_\mu + \Delta_\mu^\perp$ \\
$\nb_\mu \to \nb_\mu -\frac{\Delta^\bot_\mu}{(n\cdot v)^2}$\\ 
$v_\mu\to v_\mu$
\end{tabular}
\right. \,, \qquad\qquad\quad
(\$) \left\{
\begin{tabular}{l}
$v^\mu \to v^\mu + \beta_T^\mu$ \\
$\nb^\mu \to \nb^\mu +\frac{2}{n\cdot v}\: \beta_\perp^\mu$ \\
$n_\mu\to n_\mu$
\end{tabular}
\right. \,,
\end{eqnarray}
where $\Delta^\bot\sim \lambda$, $\beta\sim \lambda^2$, and $\beta_\perp$ is the
$\perp$-part of $\beta_T$. In defining the \$-transformation we found that it is
more convenient to leave $v_\perp=0$ by making a transformation on $v$
simultaneously with $\bn$, rather than simultaneously with $n$.  Under the
$\star$-transformation the components of a generic four-vector $V_\mu$ transform
as
\begin{eqnarray}
 n \cdot V && \stackrel{\star}{\longrightarrow} n
    \cdot V + \Delta^\perp \cdot V^\perp \,, \nn\\
 \bn \cdot V && \stackrel{\star}{\longrightarrow} \bn \cdot V -
     \frac{\Delta^\perp \cdot V^\perp}{(\ndv)^2} \,, \nn\\
V^\perp_\mu && \stackrel{\star}{\longrightarrow} V^\perp_\mu + \Delta^\perp_\mu
    \Big(-\frac{\bn}{2} +\frac{n}{2(\ndv)^2}\Big)\cdot V + \Big(-\frac{\bn_\mu}{2}
     +\frac{n_\mu}{2(\ndv)^2}\Big)\Delta^\perp\cdot V^\perp \,.
\end{eqnarray}
To the order we are working we need the following terms from an RPI-$\star$
transformation: 
\begin{eqnarray}
\delta_\star^{(\lambda^0)} {\mathcal P}_\perp^\mu &=&  -\frac{\Delta_\perp^\mu}{2}\: \bnP 
    \,,\qquad\qquad\qquad \ \ \:
\delta_\star^{(\lambda^0)} (ig n \mcdot {\mathcal B}) = 
    \Delta_\perp \cdot (ig {\mathcal B_\perp}) 
    \,, \\
\delta_\star^{(\lambda^0)}(ig {\mathcal B_\perp^\mu})  &=&  0
    \,,\qquad\qquad\qquad\qquad\qquad\quad \ \ \, 
\delta_\star^{(\lambda^0)} \bar{\chi}_n = 0
    \,,\nn \\
 \delta_\star^{(\lambda^1)} {\mathcal P}_\perp^\mu &=&
     (\bn\mcdot v)^2 n_T^\mu\:
    %\frac12\: {[n^\mu (\bn\mcdot v)^2-\bn^\mu]} \:
   \Delta^\perp \mcdot {\mathcal P}_\perp 
    \,,\qquad\qquad\quad  
\delta_\star^{(\lambda^1)}\bar{\chi}_n = \bar{\chi}_n \:\frac{\bnslash
  \Delta\!\!\!\!\slash^\perp}{4} 
  \,, \nn \\[4pt]
\delta_\star^{(\lambda^1)} (ig{\mathcal B^\mu_\perp}) &=& 
  (\bn\mcdot v)^2 n_T^\mu\:
 %\frac12 \: {[n^\mu (\bn\mcdot v)^2-\bn^\mu]} \:
  \Delta^\perp \mcdot (ig{\mathcal B}_\perp)
  \,,\qquad\qquad\!\!
 \delta_\star^{(\lambda^2)} \bnP =
    - (\bn\mcdot v)^2 \Delta_\perp\cdot {\mathcal P}_\perp 
    \,,\nn \\[6pt]
\delta_\star^{(\lambda^2)}\bar{\chi}_n &=& 
  -\bar{\chi}_n \Big( ig {\mathcal B}\!\!\!\slash^\perp + 
  {\mathcal P}\!\!\!\!\slash^{\perp \, \dagger} \Big) 
  \frac{1}{{\bar{\mathcal P}}^\dagger} \frac{\Delta\!\!\!\!\slash^\perp}{2 (\ndv)^2}
   + \bar{\chi}_n \Big[ \frac{1}{(\ndv)^2\bar{{\mathcal P}}}\:
   ig {\mathcal B}^\perp \mcdot \Delta^\perp \Big]
  \,,\nn 
\end{eqnarray}
where $n_T$ is the transverse part of $n$,
\begin{eqnarray}
    n_T^ \mu = n^\mu-\ndv\, v^\mu 
             = \frac{n^\mu}{2} -{(\ndv)}^2 \frac{\nb^\mu}{2} \,.
\end{eqnarray} 
We will also need the transformation
\begin{eqnarray}
\delta_\star^{(\lambda^2)} \delta(\omega-n\mcdot v\bnP^\dagger) &=& 
  \frac{1}{n\mcdot v} \: 
   \Delta_\perp\mcdot  {\mathcal P}_\perp^\dagger \:
  \delta'(\omega-n\mcdot v \bnP^\dagger) 
  \,.
\end{eqnarray}

For the RPI-$\$$ transformation at the order we are working we need the
following terms:
\begin{align}
 \delta_\$^{(\lambda^0)} {\mathcal H}_v &= (i m \beta_T \cdot x ) {\mathcal H}_v 
   \,,
 &\delta_\$^{(\lambda^2)} {\mathcal H}_v &= 
    \frac{\beta\!\!\!\slash_T}{2}\  {\mathcal H}_v  \,, \\
 \delta_\$^{(\lambda^2)}\: \delta(\omega-n\mcdot v\bnP^\dagger) &=
   - n\mcdot \beta_T \: \bnP^\dagger \: 
   \delta'(\omega-n\mcdot v\bnP^\dagger) \,.\nn
\end{align} 
For the last identity it is straightforward to see that the \$-transformation on
$\bn$ does not enter until one higher order. We chose to define the RPI-$\$$
transformation to be for $v$ and $\bn$ rather than $v$ and $n$ because of the
property that terms with $\bn$ are often pushed to higher order, making the
relations derived with RPI-$\$$
more orthogonal to those from RPI-$\star$. For
example, in order to have a simple form for the $\delta_\$\Gamma$'s in
Eq.~(\ref{G0dollar}) below it is important that it is $\bn$ and not $n$ that
transforms. Finally, we note that since all Dirac structures are $\mathcal O
(1)$, all RPI transformations of Dirac structures have the same power counting
as the transformation parameter, in particular, $\delta_\star \Gamma \sim
\mathcal O (\lambda^1)$ and $\delta_\$ \Gamma \sim \mathcal O (\lambda^2)$.

Finally, note that we will consider the RPI tranformations of all currents prior
to making the field redefinition in Eq.~(\ref{fd}) so that we do not have to
transform $Y$. However, in order not to have to switch our notation back and
forth we will write all equations with the operators obtained after the field
redefinition. This implies that results quoted for the transformation of objects
involving ${\cal H}_v$ should be though of as being made for $h_v$, with the
field redefinition which induces ${\cal H}_v$ made only afterwards.

\subsection{Completeness of Projected RPI} \label{sect_complete}

\begin{figure} 
\begin{center}
\includegraphics[height=5cm]{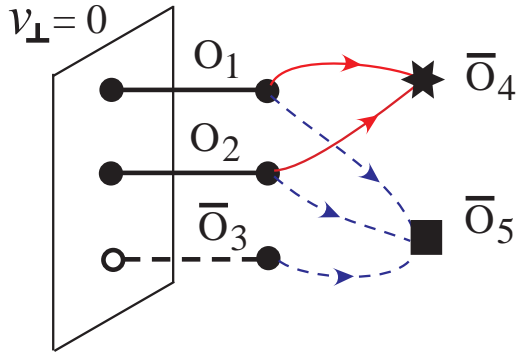} 
\end{center}
\vskip-18pt
\caption[1]{
  Transformation of operators on and off the $v_\perp=0$ surface. Here ${\mathrm
    O}_{1,2}$ exist for $v_\perp=0$, while $\overline {\mathrm
    O}_{3,4,5}$ vanish on the $v_\perp=0$ surface.
\label{fig_ops}}
\end{figure}

It is natural to ask if for $v_\perp=0$ the transformations RPI-$\$$
and
RPI-$\star$ in Eq.~(\ref{repinvnew}) are sufficient to give the complete set of
constraints that arise from the original SCET type-I, SCET type-II, and HQET RPI
transformations. The set $\{$ RPI-$\$$, RPI-$\star$, SCET-II $\}$ forms an
equivalent complete grouping related by linear combinations.  To address this
question, consider splitting all possible operators into two sets, a set $\{
{\mathrm O}_i \}$ which do not vanish on the $v_\perp=0$ surface and a set $\{
\overline {\mathrm O}_i \}$ which do. An example is pictured in
Fig.~\ref{fig_ops}.

Constraints are derived by requiring cancellations among the resulting
post-transformation set of operators. If we consider an operator $\mathrm O_i$
then under one of the projected RPI transformations, RPI-$\$$
or RPI-$\star$, it
transforms into the set $\{ \mathrm O_j , \overline {\mathrm{O}}_k \}$. On the
other hand an operator $\overline {\mathrm O}_i$ only transforms back into the
set $\{ \overline {\mathrm{O}}_j \}$. This is a special feature of the projected
transformations and ensures that relations derived on the $v_\perp=0$ surface
can not be spoiled by operators which appear away from the surface.  It appears
that we can neglect the $\overline {\mathrm O}_i$ operators since they vanish
when we project on the $v_\perp=0$ plane. However it is still possible that we
will miss an additional relation between operators on the surface, so that the
surface analysis will not be complete. 

There are two possible sources that could lead to additional relations beyond
those derived from projected RPI on the surface. First, under the SCET RPI-II
transformation $\epsilon_\perp\sim \lambda^0$ is allowed, while in the
RPI-$\star$ and RPI-$\$$
transformations we only have smaller transformations of
$\bn$ of ${\cal O}(\lambda^{1})$ and ${\cal O}(\lambda^{2})$.  Thus we could
miss relations from the more restrictive $\epsilon_\perp\sim \lambda^0$ allowed
by SCET RPI-II. Note that an SCET RPI-II transformation takes us off the
projected surface.  Second if we project onto $v_\perp=0$ then constraints are
derived only by enforcing cancellations within the set $\{ \mathrm O_j \}$. It
is possible that an operator ${\overline {\mathrm O}}_4$ exists that is obtained
from the transformation of two operators $O_1$ and $O_2$ that are {\em not
  related by transformations on the surface}.  Enforcing the cancellation of
${\overline {\mathrm O}}_4$ then relates ${\mathrm O_1}$ and ${\mathrm O_2}$.
This is pictured by the star in Fig.~\ref{fig_ops}.  A related alternative is an
operator like ${\overline {\mathrm O}}_5$ pictured with the box which is
obtained from transformations of ${\mathrm O_{1,2}}$ and $\overline {\mathrm
  O_3}$. If $\overline {\mathrm O_3}$ is otherwise constrained then this would
also constrain ${\mathrm O_{1,2}}$.  In cases with multiple operators appearing
and multiple transformations we must of course consider the linear independence
of combinations of operators. If an $\overline {\mathrm O}_i$ contributes and it
is not otherwise constrained then this is not of concern, since in the end we
discard ${\overline {\mathrm O}}_i$ by projecting onto the $v_\perp=0$ surface
anyway.  We will call an operator that vanishes for $v_\perp=0$ but that
generates a relation between operators on the surface a ``supplementary
projected operator'' (SPO).\footnote{In the case of type-II transformations,
  operators like $\overline {\mathrm O_{4}}$ and $\overline {\mathrm O_{5}}$
  need not be in the $\{\overline{\mathrm O}_j\}$ class.} To check for the
existence of an SPO we might in general need the full set of $v_\perp\ne 0$
operators.  At ${\cal O}(\lambda)$ the comparison of the results derived in
Ref.~\cite{Pirjol:2002km} in the full space, to those derived in
Ref.~\cite{Hill:2004if} on the surface $v_\perp=0$ shows that there are no SPO's
at this order.

For the ${\cal O}(\lambda^2)$ heavy-to-light operators considered here we show
that there also no SPO's in section~\ref{sect_constD}. This is done by a careful
choice of our Dirac basis which makes it simpler to demonstrate that there are
no further type-II RPI relations, and by explicit construction for other
possible SPO's.  Thus, the analysis on the $v_\perp=0$ surface is complete for
our computation.

%%%%%%%%%%%%%%%%%%%%%%%%%%%%%%%%%%%%%%%%%%%%%%%%%%%%%%%%%%%%%%%%%%%%%%%%
\section{Heavy-to-Light currents to ${\mathcal O}(\lambda^2)$}
%%%%%%%%%%%%%%%%%%%%%%%%%%%%%%%%%%%%%%%%%%%%%%%%%%%%%%%%%%%%%%%%%%%%%%%%

To order $\lambda^2$, the operators and Wilson coefficients for the
heavy-to-light currents can be written as
\begin{eqnarray}
 J = && J^{(0)} + J^{(1)} + J^{(2)} \\
   =&& \sum_j \int\!\! d\omega\, C_j(\omega,m,\mu) J_j^{(0)}(\omega,\mu)
   + \sum_{x,j} \int\! [d\omega_i]\, B_{xj}(\omega_i,m,\mu)
   J_j^{(1x)}(\omega_i,\mu)
   \nn\\
 &&+ \sum_{x,j} \int\! [d\omega_i]\, A_{xj}(\omega_i,m,\mu)
   J_j^{(2x)}(\omega_i,\mu), \nn
\end{eqnarray}
where $J^{(kx)}(\omega_i)$ represents the $\O(\lambda^k)$ terms with dependence
on convolution parameters $\omega_i$. Here the subscript $x$ distinguishes
distinct field structures at a given order, and $j$ sums over distinct
Dirac structures. At ${\mathcal O}(\lambda)$ we know that there are at most two
relevant convolution parameters $i=1,2$, while we will see below that at
${\mathcal O}(\lambda^2)$ there are at most three.  We will consider both
scalar, vector, and tensor currents (and the simple extension to the
pseudoscalar and axial vector cases). When necessary we add an $(s)$, $(v)$, or
$(t)$ superscript to the Wilson coefficients in order to distinguish these
cases, e.g. $B_{a1}^{(v)}$.

We begin in section~\ref{sect_constA} by constructing all consistent field
structures for the NNLO currents.  In section~\ref{sect_constB} we use
reparameterization invariance to derive the constraint equations for these
currents under different types of RPI invariance on the $v_\perp=0$ surface.  In
section~\ref{sect_constC} we solve the constraint equations to find the allowed
Dirac structures and obtain relations among the Wilson coefficients.  Finally,
in section~\ref{sect_constD} we show that the results from the $v_\perp=0$
surface are equivalent to those obtained if all relations in the full space were
projected onto this plane.

%%%%%%%%%%%%%%%%%%%%%%%%%%%%%%%%%%%%%%%%%%%%%%%%%%%%%%%%%%%%%%%%%%%
\subsection{Current field structures at  ${\mathcal O}(\lambda^2)$}
\label{sect_constA}

We first construct a basis of currents that is consistent with gauge invariance
and power counting and eliminate structures that are redundant by the equations
of motion and Bianchi identity. At LO and NLO the currents are
\begin{eqnarray} \label{JLO}
     J^{(0)}(\omega) &=& \cnW_{n,\omega}\Gamma {\mathcal H}_v\,, \\[5pt]
%\end{equation}
%\begin{eqnarray}
J^{(1a)}(\omega) &=& \frac{1}{\omega}\: \cnW_{n,\omega}\ppda
    \Theta^{\alpha}\sub{a}{\mathcal H}_v \nn \\[4pt]
 J^{(1b)}(\omega_{1,2}) &=&
    \frac{1}{m}\: \cnW_{n,\omega_1}
      (i g \cB^{\bot}_{\alpha})_{\omega_2}
    \Theta^{\alpha}\sub{b}{\mathcal H}_v \,. \nn
\end{eqnarray}
At NNLO we find that a convenient basis for the set of field structures for the
bilinear quark operators is
\begin{align} \label{JNNLO}
    J^{(2a)}(\omega) &= \frac{1}{2m}\: \cnW_{n,\omega}
    \Upsilon^{\sigma}\sub{a} i {\mathcal D}^{T}_{us \,\, \sigma}{\mathcal H}_v
   \,, \\[4pt]
    J^{(2b)}(\omega) &= -\frac{\ndv}{\omega}\: \cnW_{n,\omega}
    \,i \nb \cdot \overleftarrow{{\mathcal D}}_{us}\Upsilon\sub{b} {\mathcal H}_v
    \,,\nonumber\\[4pt]
    J^{(2c)}(\omega) &= -\frac{1}{\omega}\: \cnW_{n,\omega}
     i \overleftarrow{{\mathcal D}}_{us\, \alpha}^{\bot} \,
     \Upsilon^{\alpha}\sub{c}{\mathcal H}_v
     \,,\nn\\[4pt]
    J^{(2d)}(\omega) &= \frac{1}{\omega^2}\: \cnW_{n,\omega}\ppda \ppdb
        \Upsilon_{(d)}^{\alpha \beta}{\mathcal H}_v
     \,,\nn\\[4pt]
    J^{(2e)}(\omega_{1,2}) &= \frac{1}{m(\omega_1+\omega_2)} \:
        \cnW_{n,\omega_{1}}
    (i g \cBpa)_{\omega_{2}} \ppdb
    \Upsilon_{(e)}^{\alpha \beta}{\mathcal H}_v
    \,,\nonumber\\[4pt]
    J^{(2f)}(\omega_{1,2}) &= \frac{\omega_2}{m(\omega_1+\omega_2)}
       \: \cnW_{n,\omega_{1}} \bigg(\frac{\pp_{\beta}}{\omega_2}
        +\frac{\ppdb}{\omega_1}\bigg) (i g \cBpa)_{\omega_{2}}
    \Upsilon_{(f)}^{\alpha \beta}{\mathcal H}_v
    \,,\nonumber\\[4pt]
    J^{(2g)}(\omega_{1,2}) &=
        \frac{1}{m\,\ndv}\: \cnW_{n,\omega_{1}}
    \Big\{ (i g n \cdot \cB)_{\omega_{2}} + 2 (ig{\mathcal B_\perp})_{\omega_2}\cdot
     {\mathcal P}_\perp^\dagger \frac{1}{\bnP^\dagger} \Big\}
    \Upsilon_{(g)}{\mathcal H}_v
    \,,\nonumber\\[4pt]
    J^{(2h)}(\omega_{1,2,3}) &=
        \frac{1}{m(\omega_2+\omega_3)} \: \cnW_{n,\omega_{1}}
    (i g \cBpb)_{\omega_{2}}(i g \cBpa)_{\omega_{3}}
    \Upsilon_{(h)}^{\alpha \beta}{\mathcal H}_v \,, \nn\\
    J^{(2i)}(\omega_{1,2,3}) &=
   \frac{1}{m(\omega_2+\omega_3)}  
   \r{Tr}[(i g \cBpb)_{\omega_{2}}(i g \cBpa)_{\omega_{3}}]   
   \: \cnW_{n,\omega_{1}}
    \Upsilon_{(i)}^{\alpha \beta}{\mathcal H}_v \,. \nn
\end{align}
%\marginpar{(2a,2b,2c,2f) same; (2e$\to$2d, 2g$\to$2e, \phantom{xxxxxxx}, 
%2h$\to$2g, 2i$\to$2h,  2d$\to$2i)} 
For a basis of four quark operators we take
\begin{align} \label{JNNLO4}
  J^{(2j)}(\omega_{1},\omega_{2},\omega_{3})&=
    \sum_{f=u,d,s}\bigl[ \cnW^f_{n,\omega_2} \Upsilon_{(j\chi)} 
    \chi_{n,\omega_3}^f\bigr]
     \bigl[\cnW_{n,\omega_1} \Upsilon_{(j\mathcal H)} {\mathcal H}_v \bigr]
    \,, \nn\\
  J^{(2k)}(\omega_{1},\omega_{2},\omega_{3})
    &=\sum_{f=u,d,s} \bigl[\cnW^f_{n,\omega_2} T^A
     \Upsilon_{(k\chi)} \chi_{n,\omega_3}^f \bigr]
    \bigl[  \cnW_{n,\omega_1} T^A \Upsilon_{(k\mathcal H)} {\mathcal H}_v\bigr]
\end{align}
where the matrices $T^A$ are generators of SU(3) with an implied sum on $A$ and
$\chi^f_n$ has a collinear quark with flavor $f$, whereas $\chi_n$ carries the
flavor of quark from the full theory current. We impose the RPI type-III
invariance in Eq.~(\ref{repinv}) on all operators by multiplying by an
appropriate power of $n\mcdot v$. The basis in
Eqs.~(\ref{JLO},\ref{JNNLO},\ref{JNNLO4}) is valid whether or not we take
$v_\perp=0$. The $v_\perp=0$ choice only effects the basis of Dirac structures.

The 11 operators in Eqs.~(\ref{JNNLO},\ref{JNNLO4}) can be compared with the 15
field structures in the basis of Ref.~\cite{Beneke:2004in}.  We have no analog
of their $J^{(2)}_{1,2,3,7}$ currents which have an explicit $x^\mu$ because
with momentum labels the multipole expansion is performed directly in momentum
space~\cite{Luke:1999kz}. Correspondingly, our $J^{(2b)}$ and $J^{(2c)}$
currents have no analogs in their basis. There is a correspondence,
$J^{(2a,2d)}\leftrightarrow J^{(2)}_{4,6}$, $\ J^{(2e,2f,2g)}\leftrightarrow
J^{(2)}_{8,9,10}$, $\ J^{(2i,2j,2k)}\leftrightarrow J^{(2)}_{13,14,15}$, and our
$J^{(2h)}$ encodes their $J^{(2)}_{12}$ and $J^{(2)}_{13}$ currents.

In arriving at Eq.~(\ref{JNNLO}) we have used Eq.~(\ref{toBD}) to switch to a
basis with ${\mathcal P}_\perp$'s, $in\mcdot \partial$, and field strengths
rather than collinear covariant derivatives in order to give simpler constraints
from RPI.  The basis with covariant derivatives is more natural from the point
of view of tree level matching and the relation between the two is discussed in
section~\ref{sect_tree}. The prefactors in $J^{(2a-2i)}$ have been chosen with
these relationships in mind, in order to make the matching coefficients for the
operators simple. The combinations in $J^{(2f,2g)}$ were chosen because they
have simpler transformations under RPI.

Structures were also removed from Eq.~(\ref{JNNLO}) using equations of motion
and the Bianchi identity.  In the effective field theory this gives a valid
basis at any loop order.  After decoupling the usoft gluons the LO Lagrangian
for collinear quarks is~\cite{Bauer:2001yt}
\begin{eqnarray}
  {\mathcal L}_c^{(0)} &=& \overline\xi_n \frac{\bnslash}{2} \Big( i n\mcdot D_c +
    i{\slash\!\!\!\! D}_c^\perp W \frac{1}{\bnP} W^\dagger 
    i{\slash\!\!\!\! D}_c^\perp \Big) \xi_n
  =\overline \chi_n \frac{\bnslash}{2} \Big( in\mcdot {\mathcal D}_c 
  + i{\slash\!\!\!\!\mathcal D}_c^\perp  \frac{1}{\bnP}
    i{\slash\!\!\!\!\mathcal D}_c^\perp \Big) \chi_n \,,
\end{eqnarray}
so the equation of motion for $\chi_n$ can be written
\begin{eqnarray} \label{eom1}
  in\mcdot \partial \chi_n = -(ig n\mcdot {\mathcal B}) \chi_n 
  - i\,{\slash\!\!\!\!\mathcal D}_c^\perp  \frac{1}{\bnP}
    i\,{\slash\!\!\!\!\mathcal D}_c^\perp \: \chi_n \,,
\end{eqnarray}
where using Eq.~(\ref{toBD}) the last term can be written as a sum of terms with
either two ${\mathcal P}_\perp$'s, two $(ig{\mathcal B}_\perp)$'s, or one of
each.  Eq.~(\ref{eom1}) shows that a a current $\bar\chi_n i n\mcdot \pleft \hc
{\mathcal H}_v $ is redundant by the collinear quark equation of motion and need
not be included in the list, explaining why we only have $J^{(2b)}$ and
$J^{(2c)}$.  (Note that $in\mcdot {\mathcal D}_{us} \chi_n = in\mcdot\partial
\chi_n$.)  As noted in~\cite{Beneke:2004in}, this makes their $J_5^{(2)}$
current redundant. In $J^{(2a)}$ we have restricted the ultrasoft derivative
acting on $\hv$ to be purely transverse since the heavy quark equation of motion
is $ v\cdot D_{us}\,\hv\eq0$.

One can also consider using the collinear gluon equation of motion.  After the
field redefinition in Eq.~(\ref{fd}), the lowest order collinear gluon Lagrangian
is the same as in QCD~\cite{Bauer:2001yt}, ${\mathcal L}_g^{(0)}= 1/(2g^2)
{\textrm{ tr}}\{ [iD_c^\mu,i D_c^\nu]\}^2$. Varying ${\mathcal
  L}_c^{(0)}+{\mathcal L}_g^{(0)}$ with respect to the collinear gluon field
$A_{c\,\mu}^{A}$ and contracting with $\nb_\mu T^A$ gives
\begin{eqnarray}
0=\nb_\mu T^A \frac{\delta \L^{(0)}}{\delta A_{c\,\mu}^{A}}
=\frac{1}{g} \nb_\mu \left[ i D_{c\nu} , \left[ i D^\mu_c , i D^\nu_c \right]  
   \right] +  g  T^A \sum_f \cn^f T^A \nbs\, \xi_n^f \,.
\end{eqnarray}
Next we multiply by $W^\dagger$ on the left and $W$ on the right, use the
identity $(W^\dagger T^A W)\otimes T^A = T^A \otimes(W T^A W^\dagger)$, and
label by $\omega_2$ to give
\begin{align}
  -g^2  T^A \sum_{f} \big[ {\bar\chi}_{n}^f\: T^A \nbs 
     \chi_{n}^f \big]_{\omega_2} &=
   \big( \left[ i {\mathcal D}_{c\nu} , \left[ i \bn\mcdot {\mathcal D}_c , 
      i {\mathcal D}^\nu_c \right]  \right] \big)_{\omega_2} \\
 &= \frac{\omega_2^2}{2} (ig n\mcdot {\mathcal B})_{\omega_2} 
     - \omega_2 {\mathcal P}^\perp_\nu ( i g {\mathcal B}_\perp^\nu)_{\omega_2} 
     - \sum_{\omega_3} \omega_3 \big[ (ig{\mathcal B}_\perp^\nu)_{\omega_2-\omega_3}, 
         (ig {\mathcal B}^\perp_{\nu})_{\omega_3} \big] \,. \nn
\end{align}
Multiplying by $\bar\chi_{n,\omega_1}$ on the left and $ \Gamma {\mathcal H}_v$
on the right where $\Gamma$ is some Dirac structure gives
\begin{align} \label{eom2}
 & \frac{\omega_2^2}{2} \: \bar\chi_{n,\omega_1} 
    (ig n\mcdot {\mathcal B})_{\omega_2} \Gamma {\mathcal H}_v 
   = - g^2 \sum_{f,\omega_3} \big[ {\bar\chi}_{n,\omega_2-\omega_3}^f T^A 
  \nbs \chi_{n,\omega_3}^f \big]
  \big[ \bar\chi_{n,\omega_1}  T^A \Gamma {\mathcal H}_v \big] \\
 & \hspace{2.5cm} + \omega_2 \: \bar\chi_{n,\omega_1} {\mathcal P}^\perp_\nu 
      ( i g {\mathcal B}_\perp^\nu)_{\omega_2} \Gamma {\mathcal H}_v 
 + \sum_{\omega_3} \omega_3\:  \bar\chi_{n,\omega_1} 
    \big[ (ig{\mathcal B}_\perp^\nu)_{\omega_2-\omega_3}, 
         (ig {\mathcal B}^\perp_{\nu})_{\omega_3} \big] \Gamma {\mathcal H}_v
    \,. \nn 
\end{align}
This result can be used to eliminate the current $J^{(2g)}(\omega_{1,2})$ in
terms of $J^{(2e)}$ and $J^{(2k)}$ if desired. We have chosen not to remove this
operator since doing so would induce a tree level matching contribution for
$J^{(2k)}$.  For listing results it was more convenient to leave all four quark
operators with coefficients that start at one-loop order, ${\mathcal
  O}(\alpha_s^2)$. Since Eq.~(\ref{eom2}) eliminates a current that will not
show up in the constraint equations it does not effect the discussion of RPI
relations.

The Bianchi identity in QCD is $D_\mu G_{\nu\sigma}+D_\nu G_{\sigma\mu} +
D_{\sigma} G_{\mu\nu}=0$. It can be used to eliminate terms proportional to $ig
B_{\perp\perp}^{\mu\nu} \equiv [ iD_{\perp}^\mu , i D_{\perp}^\nu ]$  or
\begin{eqnarray}
  i g {\cal B}_{\perp\perp}^{\mu\nu} \equiv \Big[ \frac{1}{\bnP} W^\dagger i g
  B_{\perp\perp}^{\mu\nu} W \Big] \,,
\end{eqnarray}
in terms of factors of $i g {\cal B}_\perp^\lambda$ or $ig B_\perp^\lambda =
[i\bn\mcdot D , iD_{\perp}^\lambda]$.  The Bianchi identity gives $[\bn\mcdot D,
B_{\perp\perp}^{\mu\nu} ]= [D_\perp^\mu, B_\perp^\nu] - [D_\perp^\nu,
B_\perp^\mu]$ so using Eq.~(\ref{toBD}) we have
\begin{eqnarray}
   (ig {\cal B}_{\perp\perp}^{\mu\nu})
    = \frac{{\cal P}_\perp^\mu}{\bnP} (ig {\cal B}_\perp^\nu) 
    - \frac{{\cal P}_\perp^\nu}{\bnP} (ig {\cal B}_\perp^\mu) 
    +\frac{1}{\bnP^2} 
      \big[ (ig {\cal B}_\perp^\mu) ,  (\bnP ig {\cal B}_\perp^\nu) \big]
    - \frac{1}{\bnP^2} 
      \big[ (ig {\cal B}_\perp^\nu) ,  (\bnP ig {\cal B}_\perp^\mu) \big] 
  \,.
\end{eqnarray}
Thus a heavy-to-light current with $(ig {\cal B}_{\perp\perp}^{\mu\nu})$ can be
matched onto a linear combination of $J^{(2f,2e)}$ and $J^{(2h)}$ with
antisymmetric indices in $\Upsilon_{(h)}^{\alpha\beta}$.

%%%%%%%%%%%%%%%%%%%%%%%%%%%%%%%%%%%%%%%%%%%%%%%%%%%%%%%%%%%%%%%%%%%%%%%%

\subsection{Constraint equations from reparameterization invariance}
\label{sect_constB}

We derive constraint equations for the allowed subleading currents considering
the different types of RPI in turn.

%%%%%%%%%%%%%%%%%%%%%%%%%%%%%%%%%%%%%%%%%%%%%%%%%%%%%%%%%%%%%%%%%%%%%%%%
\subsubsection{RPI-$\star$ at ${\mathcal O}(\lambda)$}

To set the stage we review the constraints at ${\mathcal O}(\lambda)$ from SCET
RPI.  To ensure that the next-to-leading order current is RPI-$\star$ invariant,
we must have
\begin{eqnarray} \label{SCETs1}
    \deltas1_\star  J^{(0)} + \deltas0_\star  J^{(1)} = 0.
\end{eqnarray}
Computing the various terms in this equation gives\footnote{Note the remark on
  our use of notation at the end of section~\ref{sect_rpi} that explains why we
  do not include the transformation of $Y$.}
\begin{align}
    \deltas1_\star  &J^{(0)}(\omega) = \cnW_{n,\omega} \big(\frac{1}{4}  
    \nbs \Deltaps \Gamma
    + \deltas1_\star  \Gamma \big) {\mathcal H}_v \,, \nonumber\\
    \deltas0_\star  &J^{(1a)}(\omega) = \frac{1}{\omega}\cnW_{n,\omega}
   \big({-\frac{1}{2}} \Deltapa \pbd \big) 
    \Theta^{\alpha}\sub{a} {\mathcal H}_v \,, \nonumber\\
    \deltas0_\star  &J^{(1b)}(\omega_{1,2}) = 0.
\end{align}
The terms that must cancel all have a common dependence on
$\bar\chi_{n,\omega}$, $\Deltapa$, and ${\mathcal H}_v$ which can be
factored out. The remaining coefficients and Dirac structures give the
constraint equation:
\begin{eqnarray}\label{constraint1}
\boxed{ 
  \sum_j B_{aj}(\omega)\Theta^{\alpha}\sub{aj} 
   = \sum_j C_j(\omega) \Big( \frac{1}{2} \Nbs
    {\gammapa} \Gamma_{(j)} + 2 \deltaa_\star  \Gamma_{(j)} \Big) 
}
\end{eqnarray} 
where the index $\alpha$ is $\perp$, $j$ sums over Dirac structures, and
$\deltaa_\star  \Gamma_{(j)}$ is defined through
\begin{equation}\label{contraint1defn} 
    \deltas{1}_\star  \Gamma_{(j)} = \frac{1}{n\mcdot v}
 \, \Deltapa \deltaa_\star  \Gamma_{(j)} \,.
\end{equation}

%%%%%%%%%%%%%%%%%%%%%%%%%%%%%%%%%%%%%%%%%%%%%%%%%%%%%%%%%%%%%%%%%%%%%%%%
\subsubsection{RPI-$\$$ at ${\mathcal O}(\lambda^2)$}

The only terms in the current whose transformation under RPI-\$ leaves
uncanceled terms are $J^{(0)}$ and $J^{(2a)}$.  We must have
\begin{equation}
\delta_\$^{(\lambda^2)}J^{(0)} + \delta_\$^{(\lambda^0)} J^{(2a)} = 0.
\end{equation}
Now,
\begin{align}
    \deltas2_\$ J^{(0)}(\omega) &=
    \bar{\chi}_n \bigl[{-n \cdot \beta_T}\,\pbd
     \delta ' (\omega- n \cdot v \pbd ) \bigr]
    \Gamma {\mathcal H}_v +
    \bar{\chi}_{n, \omega} \Big[
  \delta_\$^{(\lambda^2)}\Gamma + \frac{\Gamma\beta\!\!\!\slash_T}{2}\: \Big]
   {\mathcal H}_v \,, \nn \\
  \deltas0_\$ J^{(2a)}(\omega) &=  \frac{1}{2 m}\bar{\chi}_{n, \omega}
    \Upsilon^\sigma\sub{a} (-m \beta^T_\sigma) {\mathcal H}_v.
\end{align}
Suppressing the common fields $\bar\chi_{n,\omega}$, ${\mathcal H}_v$, and
vector $\beta^\sigma_T$ leads to the constraint equation
\begin{equation} \label{constrainth}
\boxed{
   \sum_j A_{aj}(\omega) \Upsilon^\sigma\sub{aj} 
   = \sum_j \bigg\{ C_j(\omega) \big(2\delta_\$^\sigma \Gamma_{(j)}
   + \Gamma_{(j)} \gamma_T^\sigma \big)
   +  2 \omega \frac{d}{d\omega}C_j(\omega)\: \frac{n_T^\sigma}{\ndv} \Gamma_{(j)} 
  \bigg\}
  \,, 
}
\end{equation} 
where
\begin{equation} 
 \delta_\$^{(\lambda^2)}\Gamma_{(j)} = \beta_\sigma^T \delta_\$^\sigma \Gamma_{(j)}
\qquad\text{and} \qquad \gamma_T^\sigma=\gamma^\sigma-\slashed{v}v^\sigma.
\end{equation}

%%%%%%%%%%%%%%%%%%%%%%%%%%%%%%%%%%%%%%%%%%%%%%%%%%%%%%%%%%%%%%%%%%%%%%%%
\subsubsection{SCET RPI-a at ${\mathcal O}(\lambda^2)$}

The terms in the current that transform under SCET RPI-a are
$J^{(0)}, J^{(1a)}, J^{(2b)},\ \text{and}\ J^{(2c)}$.  We must have
\begin{equation}
    \deltas2_a J^{(0)} + \deltas1_a J^{(1a)} + \deltas0_a J^{(2b)} 
     + \deltas0_a J^{(2c)} = 0.
\end{equation}
Now,
\begin{align}
    \deltas2_a J^{(0)}(\omega) &= \cnW_n \bigl[{-\ndv\,\bn \cdot \ell} 
    \, \delta' (\omega- n \mcdot v \pbd )\bigr] \Gamma\, {\mathcal H}_v
    \,,\nn\\[4pt]
    \deltas1_a J^{(1a)}(\omega) &= \frac{1}{\omega}\:
    \cnW_{n,\omega} \, \ell^\perp_\alpha \,
    \Theta^{\alpha}\sub{a}{\mathcal H}_v 
    \,,\nn\\[4pt]
    \deltas0_a J^{(2b)}(\omega) &= -\frac{\ndv}{\omega}\: \cnW_{n,\omega}
    \Upsilon\sub{b} \, \nb \cdot \ell \, {\mathcal H}_v
    \,,\nn\\[4pt]
    \deltas0_a J^{(2c)}(\omega) &= -\frac{1}{\omega}\: \cnW_{n,\omega}
    \Upsilon^{\alpha}\sub{c} \, \ell_\alpha^{\bot}\, {\mathcal
    H}_v \,.
\end{align}
This leads to the a constraint equation between ${\mathcal O}(\lambda^2)$ and
${\mathcal O}(\lambda^0)$
\begin{equation} \label{constrainta1}
\boxed{
  \sum_j A_{bj}(\omega) \Upsilon_{(bj)}=  \sum_j \omega C'_j(\omega) \Gamma_{(j)}
}
\end{equation} 
and a constraint equation between ${\mathcal O}(\lambda^2)$ and ${\mathcal
  O}(\lambda)$
\begin{equation} \label{constrainta2}
\boxed{
 \sum_j A_{cj}(\omega)\Upsilon_{(cj)}^\alpha = \sum_j B_{aj}(\omega) \Theta_{(aj)}^\alpha \,.
} 
\end{equation}

%%%%%%%%%%%%%%%%%%%%%%%%%%%%%%%%%%%%%%%%%%%%%%%%%%%%%%%%%%%%%%%%%%%%%%%%
\subsubsection{SCET RPI-$\star$ at ${\mathcal O}(\lambda^2)$} 

Under RPI-$\star$ we must have
\begin{equation}
    \deltas2 J^{(0)} + \deltas1 J^{(1)} + \deltas0 J^{(2)} = 0.
\end{equation}
Many of the currents transform under this form of RPI:
\begin{align} \label{scetS2}
\deltas{2}_\star J^{(0)}(\omega) &= -\cnW_n
  \bigg\{
      i g \cBps \Deltaps \frac{(\nbdv)^2}{2\pbd}
      \!+\! \Big[ i g \cBp \mcdot \Deltap \frac{(\nbdv)^2}{\pbd} \Big]
   %         \nn\\ 
   %        & \mspace{120mu} 
     + \ppds \Deltaps \frac{(\nbdv)^2}{2\pbd}
     \Big\}       \delta(\omega\!-\!\ndv \pbd)
      \nn\\
  &  \mspace{75mu}
     \! -\!  \Deltap \cdot \ppd  {(\nbdv)}\:
      \delta' (\omega \!-\! n \cdot v \pbd )  
     \bigg\} \Gamma \, {\mathcal H}_v
    \,, \nonumber\\[4pt]
\deltas{1}_\star J^{(1a)}(\omega) &= \frac{1}{\omega}\: \cnW_{n,\omega}
        \bigg\{ \ppda \Bigl(\frac{\nbs \Deltaps}{4}  \Theta^{\alpha}\sub{a}
        + \deltas1_\star  \Theta^{\alpha}\sub{a} \Bigr)
%        \nn\\
%  &\mspace{219mu}
        + (\nbdv)^2 \Deltap \mcdot \ppd n^{T}_{\sigma}
         \Theta^{\sigma}\sub{a} \bigg\} {\mathcal H}_v
    \,,\nonumber\\[4pt]
\deltas{1}_\star J^{(1b)}(\omega_{1,2}) &=\frac{1}{m}
       \cnW_{n,\omega_{1}} \bigg\{
        i g \cBpa\, \Big( \frac{\nbs\Deltaps}{4}   \Theta^{\alpha}\sub{b}
        + \deltas1_\star  \Theta^{\alpha}\sub{b}  \Big)
%        \nn\\
% &\mspace{140mu}
        +(\nbdv)^2\,\Deltap\mcdot (ig\cBp) \, n^{T}_\sigma 
         \Theta^\sigma\sub{b}
         \bigg\}_{\omega_{2}}\!\! {\mathcal H}_v
        \,,\nonumber\\[4pt]
  & \mspace{-100mu}  \deltas0_\star  J^{(2a,2b,2c)} = 0 \,,\nonumber\\
\deltas0_\star  J^{(2d)}(\omega) &= \frac{(\bn\mcdot v)}{\omega}\: 
    \cnW_{n,\omega}\Big( {-\frac{1}{2}} \Deltapa \ppdb
    - \frac{1}{2} \ppda \Deltapb \Big) {\mathcal H}_v 
     \,,\nonumber\\[4pt]
\deltas0_\star  J^{(2e)}(\omega_{1,2}) &=\frac{1}{m(\omega_1+\omega_2)}\:
     \cnW_{n,\omega_{1}}(i g \cBpa)_{\omega_{2}}
    \Big({ -\frac{1}{2}} \Deltapb \pbd \Big) 
    \Upsilon^{\alpha \beta}\sub{g} \: {\mathcal H}_v 
    \,,\nonumber\\[4pt]
  & \mspace{-100mu}  \deltas0_\star  J^{(2f,2g,2h,2i,2j,2k)} = 0 \,.
\end{align}
The terms in Eq.~(\ref{scetS2}) can be grouped into two unique field structures,
$[\bar\chi_{n,\omega} \Delta^\perp_\alpha {\mathcal P}_\beta^\dagger \cdots
{\mathcal H}_v]$ and $[\bar\chi_{n,\omega} \Delta^\perp_\alpha {\mathcal
  B}_\beta^\dagger \cdots {\mathcal H}_v]$, which must cancel independently. This
gives two constraint equations. The terms proportional to $\Deltapa \ppdb$ give
\begin{equation}\l{constraintp2}
\begin{split}
   \sum_j  A_{dj}(\omega)\bigl(\Upsilon^{\alpha
       \beta}\sub{dj}+\Upsilon^{\beta\alpha}\sub{dj}\bigr) &
    = \sum_j\Big\{ -C_j(\omega) \gammapb \gammapa \Gamma_{(j)}
    - 2\omega C'_j(\omega) \gpab \Gamma_{(j)} \Big\} \\
    &+ \sum_j B_{aj}(\omega)\Big(
    \frac{1}{2} \Nbs \gammapa \Theta^{\beta}\sub{aj}
    + 2\deltaa_\star  \Theta^{\beta}\sub{aj}     
    + 2\gpab \frac{n^T_\sigma}{\ndv} \Theta^\sigma\sub{aj}
\Big) .
\end{split}
\end{equation}
From Eq.~(\ref{constraint1}) we know that the index $\sigma$ on
$\Theta_{(aj)}^\sigma$ must be $\perp$ so the last term vanishes. Inserting
Eq.~(\ref{constraint1}) also simplifies the nonvanishing terms. Finally we know
that $\Upsilon_{dj}^{\alpha\beta}$ is symmetric in $\alpha$ and $\beta$. With
these simplifications we have the constraint equation
\begin{equation}\l{constraint2}
 \boxed{
\begin{split}
   \sum_j  A_{dj}(\omega) \:\Upsilon^{\alpha
       \beta}\sub{dj} &
    = \sum_j\Big\{ - \frac12 C_j(\omega) \gammapb \gammapa \Gamma_{(j)}
    - \omega C_j'(\omega) \gpab \Gamma_{(j)} 
    + \frac{1}{2} C_j(\omega) \Nbs \gammapa \delta_\star^\beta \Gamma_{(j)} \Big\} \\
    &\quad
   + \sum_j B_{aj}(\omega)\: \deltaa_\star  \Theta^{\beta}\sub{aj}     
   \, .
\end{split}
}
\end{equation}
Since the LHS is symmetric in $\alpha\beta$, all terms on the RHS that are not
symmetric should cancel.  The terms from Eq.~(\ref{scetS2}) that are
proportional to $\Deltapa \cBpb$ give another constraint
\begin{equation}\l{constraint3}
\boxed{
\begin{split}
 & \sum_j A_{ej}(\omega_1,\omega_2) \Upsilon^{\beta \alpha}\sub{ej}
    = -\sum_j C_j(\omega_1 + \omega_2)
    \Big(
        \frac{m}{\omega_1 \!+\! \omega_2} \gammapb \gammapa
        + \frac{2m}{\omega_2} \gpab
    \Big) \Gamma_{(j)} \\
    &\phantom{=}\hspace{3cm}
    + \sum_j B_{bj}(\omega_1,\omega_2)
    \Big(
        \frac{1}{2} \Nbs \gammapa \Theta^{\beta}\sub{bj}
        + 2\deltaa_\star  \Theta^{\beta}\sub{bj}
        + 2\gpab \frac{n^T_\sigma}{\ndv} \Theta^\sigma\sub{bj}
   \Big).
\end{split}
}
\end{equation}
In Eqs.~(\ref{constraint2}) and (\ref{constraint3}), the indices $\alpha$ and
$\beta$ are purely perpendicular.  The equation that defines $\delta_\star^\alpha
\Theta^\beta$ is the same as Eq.~(\ref{contraint1defn}), just with the
$\Theta^\beta$ Dirac structures.

%%%%%%%%%%%%%%%%%%%%%%%%%%%%%%%%%%%%%%%%%%%%%%%%%%%%%%%%%%%%%%%%%%%%%%%%
\subsection{Solutions to the constraint equations}
\label{sect_constC}

We now find solutions for the ${\mathcal O}(\lambda^2)$ constraints in
Eqs.~(\ref{constrainth},\ref{constrainta1},\ref{constrainta2},\ref{constraint2},\ref{constraint3}).
Note that by careful construction of our operator basis we have ensured that
each equation gives a constraint on a different NNLO operator.  

Eqs.~(\ref{constraint1},\ref{constrainth},\ref{constraint2},\ref{constraint3})
have implicit spinor indices, one or two vector indices, and a sum in $j$ over
independent structures. Since all of the equations appear between [$\bar\xi_n
\cdots {\mathcal H}_v$] they are only valid when the spinor indices are projected
onto a 4-dimensional subspace, rather than the full 16-dimensional space of
Dirac structures.

It is useful to exploit the following method to determine how many independent
Dirac structures we should have for each operator. Start by consider the three
minimal structures that appear in the trace reduction formula,
Eq.~(\ref{reduce1}), namely $\{ 1,\gamma_5,\gamma_\perp^\alpha\}$. Next for each
case write down all possible scalar objects ($v^\mu$, $n^\mu$, $g^{\mu\nu}$,
$\ldots$) to saturate the Lorentz vector indices coming from derivatives in the
operator and current indices, taking into account any symmetries. To satisfy
parity and time reversal with $\gamma_5$, we will need to have an
$\epsilon$-tensor, such as $i\epsilon_\perp^{\mu\nu} \gamma_5$. As long as the
scalar objects are linearly independent these steps give a complete basis.

At ${\mathcal O}(\lambda^0)$, a complete basis of Dirac structures for scalar,
vector, and tensor heavy-to-light currents is~\cite{Bauer:2000yr}
\begin{align} \label{G0}
 \Gamma\sub{1} &=1 \,,\quad
 \Gamma^\mu\sub{1-3} = \Big\{ \gamma^{\mu} , v^{\mu} , 
    \Nm  \Big\} \,,  \quad
\Gamma^{\mu \nu}\sub{1-4} =  \Big\{ i \sigma^{\mu \nu},
    \gamma^{[ \mu} v^{\nu ]},
    \frac{1}{\ndv} \gamma^{[ \mu} n^{\nu ] } ,
    \frac{1}{\ndv} n^{[ \mu} v^{\nu ]} \Big \}.
\end{align}
At ${\mathcal O}(\lambda)$, there is no constraint on $J^{(1b)}$, and
Eq.~(\ref{constraint1}) constrains the $J^{(1a)}$ currents in terms of
$J^{(0)}$. To impose this constraint we need
\begin{align} \label{G0star}
 \delta_\star^\alpha\Gamma_{(1)} &= 0 \,,
 & \delta_\star^\alpha\Gamma^\mu_{(1,2)} &= 0 \,,
 & \delta_\star^\alpha\Gamma^\mu_{(3)} &= g_\bot^{\alpha\mu} \,,\\
\delta_\star^\alpha\Gamma^{\mu\nu}_{(1,2)} &= 0 \,,
 &\delta_\star^\alpha\Gamma^{\mu\nu}_{(3)} &= \gamma^{[\mu}g_\bot^{\nu]\alpha }
   \,,
 &\delta_\star^\alpha\Gamma^{\mu\nu}_{(4)} &= g_\bot^{\alpha [\mu }v^{\nu]} 
  \,. \nn
\end{align}
The constraint equation causes some Dirac structures to always appear in the
same combination. We find
\begin{eqnarray} \label{G1}
 \hspace{0cm}  \Theta^{\alpha}\sub{a1} 
  &=& \frac{1}{2} \Nbs \gammapa \,, \qquad
  \Theta^{\alpha}\sub{b1} =  \gammapa
   \,, \nn\\[4pt]
 \Theta^{\alpha \mu}\sub{a1-3} &=&
  \left \{ \frac{1}{2} \Nbs
  \gammapa \{ \gamma^{\mu} , v^{\mu}  \}, \frac{\bnslash}{2}\,
  \gammapa n^\mu 
    +2 \gpam \right \} \,,\nn\\ 
 \Theta^{\alpha \mu}\sub{b1-4} &=&
  \left \{ %\frac{1}{2} \Nbs
  \gammapa \{ \gamma^{\mu} , v^{\mu} , \Nm  \}
    , \gpam \right \} 
    ,  \hspace{0cm} \nn \\[4pt]
\hspace{0cm}
\Theta^{\alpha \mu \nu}\sub{a1-4} &=& 
 \Big\{ \frac{1}{2} \Nbs \gammapa \{ i\sigma^{\mu\nu}, 
    \gamma^{[\mu} v^{\nu]} \} \ , \ 
    \frac{\bnslash}{2}\gamma_\perp^\alpha \gamma^{[ \mu} n^{\nu ] }
        - 2  g^{\alpha [ \mu}_{\bot} \gamma^{\nu ]}\ ,\ 
    \frac{\bnslash}{2}\gamma_\perp^\alpha n^{[ \mu} v^{\nu ]} 
        + 2 g^{\alpha [ \mu}_{\bot} v^{\nu ]}
     \Big\} \,,\nn\\
\Theta^{\alpha \mu \nu}\sub{b1-6} &=& 
 \Big\{ %\frac{1}{2} \Nbs
    \gammapa \Gamma^{\mu \nu}\sub{1-4}\ ,\
    g^{\alpha [ \mu}_{\bot} \gamma^{\nu ]}\ ,\
    g^{\alpha [ \mu}_{\bot} v^{\nu ]} \Big\} 
\,, \hspace{-.5cm}
\end{eqnarray}
where $\Gamma^{\mu\nu}\sub{1-4}$ are given in Eq.~(\ref{G0}), which is in
agreement with Ref.~\cite{Pirjol:2002km}.  This basis is equivalent to the one
in Ref.~\cite{Pirjol:2002km}.\footnote{Note that a structure $g^{\alpha [
    \mu}_{\bot} n^{\nu ]}$ is redundant in
  $4$-dimensions~\cite{Beneke:2004rc,Hill:2004if}.}  We take $\Theta_{bj}$ terms
with no $\bnslash$ so that this choice does not need to be modified if we
enlarge the basis for $v_\perp\ne 0$ (see section~\ref{sect_constD}). With
Eq.~(\ref{G1}), the constraint Eq.~(\ref{constraint1}) gives relations for the
Wilson coefficients in the $J^{(1a)}$ current
\begin{align} \label{sol1}
  B^{(s)}_{a1}(\omega)&=C^{(s)}_{1}(\omega) ,
 &B^{(v)}_{a1-3}(\omega)&=C^{(v)}_{1-3}(\omega) ,
 %&B^{(v)}_{a4}(\omega) &=2 C^{(v)}_{3}(\omega), \\[4pt]
%
 &B^{(t)}_{a1-4}(\omega)&=C^{(t)}_{1-4}(\omega) \,.
  %&B^{(t)}_{a5}(\omega)&=-2C^{(t)}_{3}(\omega) ,
  %&B^{(t)}_{a6}(\omega)&=2C^{(t)}_{4}(\omega). \nn
\end{align}
These results agree with Refs.~\cite{Chay:2002vy,Beneke:2002ph,Pirjol:2002km}.

At ${\mathcal O}(\lambda^2)$ we must solve
Eqs.~(\ref{constrainth},\ref{constrainta1},\ref{constrainta2},\ref{constraint2},\ref{constraint3}).
From these equations we see that the currents $J^{(2f)}$, $J^{(2g)}$,
$J^{(2h)}$, $J^{(2j)}$, and $J^{(2k)}$ are not constrained. The currents
$J^{(2a)}$, $J^{(2b)}$, $J^{(2c)}$, and $J^{(2d)}$ are all related to the
leading order current $J^{(0)}$. Finally the currents $J^{(2e)}$ are related to
the currents $J^{(0)}$ and $J^{(1b)}$.

To solve the equations we will need
\begin{align} \label{G0dollar}
\delta_\$^\sigma \Gamma_{(1)} &= 0 \,,
 &\delta_\$^\sigma \Gamma_{(1)}^{\mu} &= 0 \,,
 &\delta_\$^\sigma \Gamma_{(2)}^{\mu} &= g_{T}^{\sigma\mu} \,,
 &\delta_\$^\sigma \Gamma_{(3)}^{\mu} &= \nb_{T}^{\sigma}n^{\mu}\\
\delta_\$^\sigma \Gamma_{(1)}^{\mu\nu} &= 0 \,,
 &\delta_\$^\sigma \Gamma_{(2)}^{\mu\nu} &=
  \gamma^{[\mu}g_{T}^{\nu]\sigma}\,,
 &\delta_\$^\sigma \Gamma_{(3)}^{\mu\nu} &=
  \nb_{T}^{\sigma}\gamma^{[\mu}n^{\nu]} \,,
 &\delta_\$^\sigma \Gamma_{(4)}^{\mu\nu} &= 
 \nb_T^{\sigma} n^{[\mu}v^{\nu]}+\frac{1}{\ndv}n^{[\mu}g_{T}^{\nu]\sigma} 
 \,,\nn
\end{align}
where $\bn_T^\sigma = -(\bn\mcdot v)^2 n_T^\sigma$. We will also need
\begin{align} \label{G1star}
\delta_\star^\alpha\Theta^{\beta}_{(b1)} &=
   - g_\perp^{\alpha\beta}\: n\mcdot v\: \frac{\bnslash}{2}\:
%\frac{\nbs\ns}{4}g_{\bot}^{\alpha\beta}-\half \gammapa\gammapb 
 \,,\\[4pt]
\delta_\star^\alpha\Theta^{\beta\mu}_{(b1,2,3)} &= 
% \Bigl(\frac{\nbs\ns}{4}g_{\bot}^{\alpha\beta}
%  \!-\! \half \gammapa\gammapb\Bigr)
 - g_\perp^{\alpha\beta}\: n\mcdot v\: \frac{\bnslash}{2}\:
 \Gamma^{\mu}_{(1,2,3)}
  +\gammapb\delta_\star^\alpha\Gamma^\mu_{(1,2,3)} \,,
 &\delta_\star^\alpha\Theta^{\beta\mu}_{(b4)} &=
    g_\bot^{\alpha\beta}\frac{n_T^{\mu}}{\ndv} 
  \,,\nn\\[4pt]
 \delta_\star^\alpha\Theta^{\beta\mu\nu}_{(b1,2,3,4)} &= 
  %\Bigl(\frac{\nbs\ns}{4}g_{\bot}^{\alpha\beta} \!-\!
  %\half \gammapa\gammapb\Bigr)
  - g_\perp^{\alpha\beta}\: n\mcdot v\: \frac{\bnslash}{2}\:
  \Gamma^{\mu\nu}_{(1,2,3,4)}
  +\gammapb\delta_\star^\alpha\Gamma^{\mu\nu}_{(1,2,3,4)} \,,
& \delta_\star^\alpha\Theta^{\beta\mu\nu}_{(b5)} &= 
  \frac{1}{\ndv}\, g_\bot^{\alpha\beta}n_T^{[\mu}\gamma^{\nu]} \,,
  \nn\\[4pt]
\delta_\star^\alpha\Theta^{\beta\mu\nu}_{(b6)} &=
\frac{1}{\ndv}\, g_\bot^{\alpha\beta}n_T^{[\mu}v^{\nu]} \,,\nn
\end{align}
where $\alpha\beta$ were projected onto $\perp$ directions. Note that
$\delta_\star^\alpha\Theta_{(aj)}$ are easily obtained from these.  The
constraints in Eqs.(\ref{constrainta1},\ref{constrainta2}) have a particularly
simple solution:
\begin{align} \label{sola}
  A_{bj}(\omega) &=\omega C'_j(\omega),
  &A_{cj}(\omega) &=B_{aj}(\omega), 
  &\Upsilon_{(bj)} &=\Gamma_{(j)}, 
  &\Upsilon_{(cj)} &=\Theta_{(aj)} \,.
\end{align}
Solutions to the other equations are slightly more involved. We present
solutions to the constraint equations for the scalar, vector, and tensor
currents in turn.

%%%%%%%%%%%%%%%%%%%%%%%%%%%%%%%%%%%%%%%%%%%%%%%%%%%%%%%%%%%%%%%%
\subsubsection{Solutions for scalar and pseudoscalar currents
  at ${\mathcal O}(\lambda^2)$}
%%%%%%%%%%%%%%%%%%%%%%%%%%%%%%%%%%%%%%%%%%%%%%%%%%%%%%%%%%%%%%%%

The RPI constraints do not effect the allowed Dirac structures for scalar
currents, so we have the complete sets
\begin{align} \label{Scalar}
 \Upsilon^\sigma \sub{a1,2} &=   
     \Big\{ \gamma^\sigma_T , \frac{n^\sigma_T}{\ndv} \Big\} 
    \,, 
  %\nn \\
% 
 & \Upsilon\sub{b1}  = \Upsilon\sub{g1} &=  1
     \,, %\nn\\
 & \Upsilon^\alpha\sub{c1} &=    
    \frac{1}{2} \Nbs \gammapa 
     \,,\nn \\[5pt]
 \Upsilon^{\alpha \beta}\sub{d1}& =   
     \gpab 
     \,, %\nn  \\
%
% &\Upsilon^{\alpha \beta}\sub{f1=g1=h1,f2=g2=h2}  &=  
 & \Upsilon^{\alpha \beta}\sub{e1,2}  =  
  \Upsilon^{\alpha \beta}\sub{f1,2}  =  
 \Upsilon^{\alpha \beta}\sub{h1,2}  &=  
 \Upsilon^{\alpha \beta}\sub{i1,2}   =  
     \big \{ 2\gpab,\gammapa \gammapb \big \} 
        \,. \hspace{-2cm}
\end{align}
For the four quark operators, there are three possible Dirac structures in the
$\bar \chi_n \cdots \chi_n$ bilinear, $\{\bnslash,
\bnslash\gamma_5,\bnslash\gamma_\perp^\alpha\}$. In performing the matching onto
SCET at a scale $\sim m_b$, the light quark masses are perturbations, and for
matching onto the ${\mathcal O}(\lambda^2)$ four quark operator we can set
$m_q=0$.  In this case, chirality rules out the $\bnslash\gamma_\perp^\alpha$
structure which connects right and left handed quarks. A complete set of
structures is therefore
\begin{align} \label{Scalar4}
\big(\Upsilon\otimes\Upsilon\big)_{(j1,j2)} &= \big(\Upsilon\otimes\Upsilon\big)_{(k1,k2)} 
   =\big\{ \Nbs\ot1 \,,\ \Nbs\gamma^5\ot \gamma^5  \big\} 
   \,.
\end{align}

To solve the RPI-\$ constraint, we insert the Dirac structures
Eqs.~(\ref{G0},\ref{G0dollar},\ref{Scalar}) into Eq.~(\ref{constrainth}).
Satisfying this constraint requires a relation on the Wilson coefficients
\begin{align}  \label{Scalarstart}
 A_{a1}^{(s)}(\omega)&=C_{1}^{(s)}(\omega) &
 A_{a2}^{(s)}(\omega)&=2 \omega C_1^{(s)\,\prime}(\omega)\,.
\end{align}
The solution for the SCET RPI-a constraint equation in (\ref{sola}) gives
\begin{align} 
 A_{b1}^{(s)}(\omega)& = \omega  C_{1}^{(s)\,\prime}(\omega)&
 A_{c1}^{(s)}(\omega) &= C_{1}^{(s)}(\omega).
\end{align}
To solve the SCET RPI-$\star$ constraints in
Eqs.~(\ref{constraint2},\ref{constraint3}), we need the additional Dirac
structures in Eqs.~(\ref{G1},\ref{G1star}). On the RHS of
Eq.~(\ref{constraint2}) we observe that all structures that were not symmetric
in $\alpha\beta$ cancel, in agreement with the symmetry of the LHS.  Solving the
equations, the relations on the Wilson coefficients are
\begin{align} \label{Scalarend}
 A_{d1}^{(s)}(\omega)&=-\omega C_1^{(s)\,\prime}(\omega) \,, \\[4pt]
 A_{e1}^{(s)}(\omega_{1,2})
  &=-\ffrac{m}{\omega_2}\,
 C_1^{(s)}(\omega_{1}\!+\!\omega_2) \,,\nn \\[4pt]
 A_{e2}^{(s)}(\omega_{1,2})&=
  -\fbrac{m}{\omega_1\! +\! \omega_2}\, C_{1}^{(s)}(\omega_1\! +\! \omega_2) 
  -B_{b1}^{(s)}(\omega_{1,2}) \,. \nn
\end{align}
The following Wilson coefficients of scalar currents are not determined by the
RPI constraints
\begin{align} \label{Scalarnot}
 & A_{f1,2}^{(s)}(\omega_{1,2}) \,, & 
 & A_{g1,2}^{(s)}(\omega_{1,2}) \,, &
 & A_{h1,2}^{(s)}(\omega_{1,2,3}) \,, &
 & A_{i1,2}^{(s)}(\omega_{1,2,3}) \,, &
 & A_{j1,2,k1,2}^{(s)}(\omega_{1,2,3}) \,.
\end{align}

Since the light quark in the full theory current retains its chirality in the
effective theory current, the results for the expansion of the pseudoscalar
current, $\bar q\gamma_5 b$, are simple to extract from those for the scalar
case, $\bar q\, b$.  The Dirac structures for pseudoscalar currents may be
obtained by multiplying Eqs.~(\ref{Scalar},\ref{Scalar4}) on the left by
$\gamma_5$ and $1\otimes\gamma^5$, respectively. The constraints on the
Wilson coefficients of these currents are then identical.

%%%%%%%%%%%%%%%%%%%%%%%%%%%%%%%%%%%%%%%%%%%%%%%%%%%%%%%%%%%%%%%%%%%%%
\subsubsection{Solutions for vector and axial-vector currents 
 at ${\mathcal O}(\lambda^2)$}
%%%%%%%%%%%%%%%%%%%%%%%%%%%%%%%%%%%%%%%%%%%%%%%%%%%%%%%%%%%%%%%%%%%%%

The analysis for the scalar current can be extended to the vector currents,
where the extra Lorentz index makes ensuring that the Dirac basis is complete
slightly more difficult. We use the method discussed in
section~\ref{sect_constC} to count the number of terms in the Dirac basis prior
to imposing the RPI constraints. For the case of $\Upsilon_a^{\sigma\mu}$ the
index $\sigma$ is transverse to $v$ and we have
\begin{align}
  1: \: \{ g_T^{\sigma\mu}, n^{\sigma}_T n^{\mu}, n^{\sigma}_T v^\mu \} 
    \,, \qquad 
  \gamma_5: \{ i\epsilon_\perp^{\sigma\mu} \} 
    \,,\qquad
  \gamma_\perp^\sigma: \{ n^\mu, v^\mu  \} \,, \qquad
  \gamma_\perp^\mu:  \{ n^\sigma_T \} \,,
\end{align}
which has seven elements.  The counting for the $\Upsilon_{b,c,g}$ cases are
straightforward. For $\Upsilon_{d}^{\alpha\beta\mu}$ the indices $\alpha\beta$
are $\perp$ and symmetric. We have
\begin{align}
  1: \: \{ g_\perp^{\alpha\beta} n^\mu, g_\perp^{\alpha\beta} v^\mu 
    \} 
    \,, \qquad 
  \gamma_5: \{ \text{--} \} 
    \,,\qquad
  \gamma_\perp^\mu: \{ g_\perp^{\alpha\beta}  \} \,, \qquad
  \gamma_\perp^{\{\alpha}:  \{ g_\perp^{\beta\} \mu} \} \,,
\end{align}
so there are four elements in the basis. Finally, for
$\Upsilon_{e,f,h,i}^{\alpha\beta\mu}$ we have
\begin{align}
  1: \: \{ g_\perp^{\alpha\beta} n^\mu, g_\perp^{\alpha\beta} v^\mu 
    \} 
    , \quad 
  \gamma_5: \{ i\epsilon_\perp^{\alpha\beta} n^\mu, 
                  i\epsilon_\perp^{\alpha\beta} v^\mu \} 
    ,\quad
  \gamma_\perp^\mu: \{ g_\perp^{\alpha\beta}  \} , \quad
  \gamma_\perp^{\alpha}:  \{ g_\perp^{\beta \mu} \} ,\quad
  \gamma_\perp^{\beta}:  \{ g_\perp^{\alpha \mu} \} ,
\end{align}
so the basis has seven elements. 

For computations, a different basis choice is slightly more convenient. The
independent Dirac structures appearing on the RHS of the constraint equations
reduce the basis for $\Upsilon_{\{a1-7\}}^{\sigma\mu}$ by one further element.
For the vector currents we find
\begin{align} \label{Vector}
\Upsilon^{\sigma\mu} \sub{a1-6}
    &= \left \{  \gamma^{\mu} \gamma_T^\sigma \,, 
     v^{\mu} \gamma_T^\sigma + 2g_T^{\sigma\mu}\,,
      \Nm \gamma_T^\sigma  \,,\,
    \{ \gamma^{\mu} , v^{\mu} , \Nm \} \frac{n^\sigma_T}{\ndv}
    \, \right \}  
   ,  \\
\Upsilon^\mu \sub{b1-3} &= \Upsilon^\mu \sub{g1-3} = 
  \left\{ \gamma^{\mu} , v^{\mu} , \Nm  \right\}
   ,\qquad\  % \nn \\
\Upsilon^\mu \sub{c1-3} = \left \{ \frac{1}{2} \Nbs \gammapa \big\{
    \gamma^{\mu} , v^{\mu}  \big\},
    \frac{\bnslash}{2}\, \gammapa n^\mu  \!+\! 2 \gpam
     \right \} , \nn\\
\Upsilon^{\alpha \beta \mu}\sub{d1-4}
    &= \left \{ \gpab \{ \gamma^{\mu} , v^{\mu} , \Nm \} \,,\,
    \frac{1}{2} \Nbs \gamma^{\{ \alpha}_{\bot} g^{\beta \} \mu}_{\bot}\right \} 
  , \nonumber\\
\Upsilon^{\alpha \beta \mu}\sub{e1-7}
    &= \left \{ 2 \gpab \{ \gamma^{\mu} , v^{\mu} , \Nm \} \,,\,
    \gammapa \gammapb \{ \gamma^{\mu} , v^{\mu} , \Nm \}
    \,,\, \frac{1}{2} \Nbs 
   \gammapb \gpam \right \} 
    , \nonumber\\
\Upsilon^{\alpha \beta \mu}\sub{f,h,i1-7}
    &= \left \{ 2 \gpab \{ \gamma^{\mu} , v^{\mu} , \Nm \} \,,\,
    \gammapa \gammapb \{ \gamma^{\mu} , v^{\mu} , \Nm \}
    \,,\, % \frac{1}{2} \Nbs 
   \gammapb \gpam \right \} 
   .\nn 
\end{align}
The index symmetrization means $\gamma^{\{ \alpha}_{\bot} g^{\beta \}
  \mu}_{\bot}=\gamma^{\alpha}_{\bot} g^{\beta \mu}_{\bot} +
\gamma^{\beta}_{\bot} g^{\alpha \mu}_{\bot}$. In Eq.~(\ref{Vector}) we have used
Eq.~(\ref{reduce1}) to remove redundant structures.

The operators $J^{(2a,2b,2c)}$ bear some similarity to the complete basis of six
$1/m$ suppressed heavy-to-light currents in HQET~\cite{Falk:1990de,Falk:1992fm}.
The differences are due to the fact that for a collinear light quark we have the
vector $n^\mu$ available to build additional structures and from the fact that
working in the $v^\perp=0$ frame, we do not need operators like $\bar \chi_n i\,
v\mcdot \overleftarrow {\mathcal D}^\perp \Upsilon {\mathcal H}_v$.

For the four quark operators, a basis of Dirac structures is
\begin{align} \label{Vector4}
\big(\Upsilon \ot \Upsilon \big)_{(j1-6)}^\mu
    &=\big(\Upsilon \ot \Upsilon \big)_{(k1-6)}^\mu=\Big\{
    \Nbs\ot \{ \gamma^{\mu} , v^{\mu} , \Nm \}\,,\  
    \Nbs \gamma^5\ot \gamma^5 \{ \gamma^{\mu} , v^{\mu} , \Nm \} \Big\}
  \,.
\end{align}
Here the counting of the number of independent structures proceeds in the same
way as for the bilinear operators, except that we start by writing down minimal
structures for the four quark operator where we impose the correct chirality on
the purely collinear fermion bilinear. For $J^{(2j)}$ we start with six
structures, $\{ \bnslash \,, \bnslash\gamma_5 \} \otimes \{ 1,\gamma_5
,\gamma_\perp^\alpha\}$, and find that only the six terms
\begin{align}
  \bnslash\otimes 1: \{ v^\mu, n^\mu\} ,\qquad
  \bnslash\otimes \gamma_\perp^\mu: \{ 1 \} ,\qquad
  \bnslash\gamma_5 \otimes \gamma_\perp^\alpha: \{ i\epsilon_\perp^{\alpha\mu}\}
    ,\qquad
  \bnslash\gamma_5 \otimes \gamma_5: \{ v^\mu , n^\mu \} \,,
\end{align}
are allowed, which we swap for the basis in Eq.~(\ref{Vector4}). The analysis of
discrete symmetries for these currents is similar to that of the four quark
operators in the HQET Lagrangian~\cite{Blok:1996iz}.

Using Eqs.~(\ref{G0},\ref{G0dollar}), the relations for the vector current
coefficients obtained by solving the RPI-\$ constraint in
Eq.~(\ref{constrainth}) are
\begin{align} \label{Vectorstart}
A_{a1-3}^{(v)}(\omega)&=C_{1-3}^{(v)}(\omega) \,,
&A_{a4}^{(v)}(\omega) &=2\omega  C^{(v)\,\prime}_1(\omega) \,, \nn\\
 A_{a5}^{(v)}(\omega) &=2 \omega C_2^{(v)\,\prime}(\omega) \,,  
& A_{a6}^{(v)}(\omega) &= 
    -2 C_3^{(v)}(\omega) \!+\! 2 \omega C^{(v)\,\prime}_3(\omega)\,, %\nn\\
%
%A_{a7}^{(v)}(\omega) &= C_2^{(v)}(\omega) \,. 
%
\end{align}
The RPI-a solution in Eq.~(\ref{sola}) gives 
\begin{align} 
 A_{b1-3}^{(v)}(\omega) &= \omega C^{(v)\,\prime}_{1-3}(\omega) \,,
& A_{c1-3}^{(v)}(\omega) &= C_{1-3}^{(v)}(\omega)\,.
%
%& A_{c4}^{(v)}(\omega) &= 2C_{3}^{(v)}(\omega) \,.
\end{align}
Using in addition Eq.~(\ref{Vector}), we find that solving
Eq.~(\ref{constraint2}) gives
\begin{align} 
  A_{d1}^{(v)}(\omega)&=-\omega C^{(v)\,\prime}_{1}(\omega) \,,
 & A_{d2}^{(v)}(\omega)&=-\omega C^{(v)\,\prime}_{2}(\omega)
     -2C_{3}^{(v)}(\omega) \,,
  \nn \\[4pt]
 A_{d3}^{(v)}(\omega)&=-\omega C^{(v)\,\prime}_{3}(\omega)
     +2C_{3}^{(v)}(\omega) \,,
 &A_{d4}^{(v)}(\omega)&= C_{3}^{(v)}(\omega)\,.
\end{align}
Finally, solving the second RPI-$\star$ constraint in Eq.~(\ref{constraint3})
gives
\begin{align} \label{Vectorend}
A_{e1}^{(v)}(\omega_{1,2}) 
&=-\Big(\ffrac{m}{\omega_{2}}\Big)C_{1}(\omega_1\plus\omega_2)
  -B_{b3}(\omega_{1,2}) 
  \,, \\[4pt]
A_{e2}^{(v)}(\omega_{1,2}) 
&=-\Big(\ffrac{m}{\omega_{2}}\Big)C_{2}(\omega_1\plus\omega_2)
  -B_{b4}(\omega_{1,2})
 \,,\nn \\[4pt]
A_{e3}^{(v)}(\omega_{1,2}) 
&=-\Big(\ffrac{m}{\omega_{2}}\Big)C_{3}(\omega_1\plus\omega_2)
  +B_{b3}(\omega_{1,2}) +B_{b4}(\omega_{1,2})
 \,,\nn\\[4pt]
A_{e4}^{(v)}(\omega_{1,2})&=-\Big(\ffrac{m}{\omega_{1}\plus\omega_2}\Big)
C_{1}(\omega_1\plus\omega_2) +B_{b1}(\omega_{1,2})+2B_{b3}(\omega_{1,2})
 \,,\nn\\[4pt]
A_{e5}^{(v)}(\omega_{1,2})&=-\Big(\ffrac{m}{\omega_{1}\plus\omega_2}\Big)
C_{2}(\omega_1\plus\omega_2) -2B_{b1}(\omega_{1,2})-B_{b2}(\omega_{1,2})
 \,,\nn\\[4pt]
A_{e6}^{(v)}(\omega_{1,2})&=-\Big(\ffrac{m}{\omega_{1}\plus\omega_2}\Big)
C_{3}(\omega_1\plus\omega_2) -3B_{b3}(\omega_{1,2})
 \,,\nn\\[4pt]
A_{e7}^{(v)}(\omega_{1,2})&= -2B_{b3}(\omega_{1,2}) + B_{b4}(\omega_{1,2}) \,.\nn
\end{align}
The following Wilson coefficients of the ${\mathcal O}(\lambda^2)$ vector currents
are not determined by the RPI constraints,
\begin{align} \label{Vectornot}
 & A_{f1-7}^{(v)}(\omega_{1,2}) \,, & 
 & A_{g1-3}^{(v)}(\omega_{1,2}) \,, &
 & A_{h1-7,i1-7}^{(v)}(\omega_{1,2,3}) \,, &
 & A_{j1-6,k1-6}^{(v)}(\omega_{1,2,3}) \,.
\end{align}

The Dirac structures for axial-vector currents which expand $\bar u \gamma_5
\gamma^\mu b$ may be obtained by multiplying the Dirac structures in Eq.~(\ref{Vector}) by
$\gamma_5$ on the left and in Eq.~(\ref{Vector4}) by $1\otimes\gamma^5$.  The
relations for their Wilson coefficients are then the same as the vector
currents.

%%%%%%%%%%%%%%%%%%%%%%%%%%%%%%%%%%%%%%%%%%%%%%%%%%%%%%%%%%%%%%%%%%%%%
\subsubsection{Solutions for tensor currents at ${\mathcal O}(\lambda^2)$}
%%%%%%%%%%%%%%%%%%%%%%%%%%%%%%%%%%%%%%%%%%%%%%%%%%%%%%%%%%%%%%%%%%%%%

The counting of the number of independent terms proceeds just as in the vector
case but now with antisymmetric indices $\mu\nu$. For $J^{(2a)}$, the index
$\sigma$ is transverse to $v$ and there are ten structures
\begin{align}
  1&: \{ v^{[\mu} n^{\nu]} n^\sigma, g_\perp^{\sigma[\mu} v^{\nu]} 
     , g_\perp^{\sigma[\mu} n^{\nu]} \} \,,\qquad
  & \gamma_5 &: \{ i\epsilon_\perp^{\mu\nu}n^\sigma,  
      i\epsilon_\perp^{\sigma[\mu}n^{\nu]},
      i\epsilon_\perp^{\sigma[\mu}v^{\nu]}
      \} \,,\nn\\
  \gamma_\perp^{[\mu} &: \{ v^{\nu]}n^\sigma, n^{\nu]}n^\sigma , 
      g_\perp^{\nu]\sigma} \} \,, 
  &\gamma_\perp^{\sigma} &: \{ v^{[\mu}n^{\nu]} \} \,.
\end{align}
The bases for $J^{(2b,2g)}$ are simple, while for $J^{(2c)}$ we have six terms
\begin{align}
 1&: \{  g_\perp^{\alpha[\mu} v^{\nu]} 
     , g_\perp^{\alpha[\mu} n^{\nu]} \} \,,%\qquad
  & \gamma_5 &: \{ i\epsilon_\perp^{\alpha[\mu}n^{\nu]},
      i\epsilon_\perp^{\alpha[\mu}v^{\nu]}
      \} \,, %\qquad
  & \gamma_\perp^{[\mu} &: \{ g_\perp^{\nu]\sigma} \} \,,% \qquad
  &\gamma_\perp^{\alpha} &: \{ v^{[\mu}n^{\nu]} \} \,.
\end{align}
We also have six terms for $J^{(2d)}$
\begin{align}
 1&: \{  g_\perp^{\alpha\beta} n^{[\mu} v^{\nu]} 
       \} \,,%\qquad
  & \gamma_5 &: \{ i\epsilon_\perp^{\mu\nu} g_\perp^{\alpha\beta} 
      \} \,, %\qquad
  & \gamma_\perp^{[\mu} &: \{ n^{\nu]} g_\perp^{\alpha\beta} ,
   v^{\nu]} g_\perp^{\alpha\beta} 
    \} \,,% \qquad
  &\gamma_\perp^{\{\alpha} &: \{ g_\perp^{\beta\}[\mu} n^{\nu]} ,
    g_\perp^{\beta\}[\mu} v^{\nu]} \} \,,
\end{align}
where the identity $g_\bot^{\alpha[\mu}\epsilon_\bot^{\nu]\beta} =
-g_\bot^{\alpha\beta}\epsilon_\bot^{\mu\nu}$ leaves only one term for
$\gamma_5$. Finally for $J^{(2e,2f,2h,2i)}$ we count ten terms
\begin{align}
 1&: \{  g_\perp^{\alpha\beta} n^{[\mu} v^{\nu]}, 
    g_\perp^{\alpha[\mu} g_\perp^{\nu]\beta}  
       \} \,,%\qquad
  & \gamma_5 &: \{ i\epsilon_\perp^{\mu\nu} g_\perp^{\alpha\beta} ,
      i\epsilon_\perp^{\alpha\beta} n^{[\mu} v^{\nu]}  \} \,, %\qquad
  & \gamma_\perp^{[\mu} &: \{ n^{\mu]} g_\perp^{\alpha\beta} ,
   v^{\mu]} g_\perp^{\alpha\beta} 
    \} \,,\nn\\
  \gamma_\perp^{\alpha} &: \{ g_\perp^{\beta[\mu} n^{\nu]} ,
    g_\perp^{\beta[\mu} v^{\nu]} \} \,, 
  & \gamma_\perp^{\beta} &: \{ g_\perp^{\alpha[\mu} n^{\nu]} ,
    g_\perp^{\alpha[\mu} v^{\nu]} \} \,.
\end{align}

Again only $J^{(2a)}$ has its basis of Dirac structures further restricted by
the RPI-\$ constraint in Eq.~(\ref{constrainth}), which reduces the basis by two
terms (since only eight linearly independent Wilson coefficients appear in
Eq.~(\ref{Tensorstart}) below).  For the complete set of Dirac structures for
tensor currents we find
\begin{align} \label{Tensor}
\Upsilon^{\sigma \mu \nu}\sub{a1-8} &= \Big\{ 
   i \sigma^{\mu \nu}\gamma_T^\sigma \ ,\
    \gamma^{[ \mu} v^{\nu ]}\gamma_T^\sigma -  2 g_T^{\sigma[\mu}\gamma^{\nu]}\ ,\ 
    \frac{1}{\ndv} \gamma^{[ \mu} n^{\nu ] } \gamma_T^\sigma\ ,\ 
    \frac{1}{\ndv} n^{[ \mu} v^{\nu ]} \gamma_T^\sigma 
       - 2 g_T^{\sigma[\mu}n^{\nu]}\ ,\nn\\
%\Gamma^{\mu \nu}_{(1-4)}  \gamma_T^\sigma,
&\qquad \ \Gamma^{\mu \nu}_{(1-4)}
  \frac{n_T^\sigma}{\ndv}
 \Big\} \ , \ \qquad\quad
% \,, \\
%
 \Upsilon^{\mu \nu}\sub{b,g1-4} = \Big\{\Gamma_{(1-4)}^{\mu\nu}\Big\}
  \,,\nn \\[4pt]
\Upsilon^{\alpha \mu \nu}\sub{c1-4} &=
 \Big\{ \frac{1}{2} \Nbs \gammapa \{ i\sigma^{\mu\nu}, 
    \gamma^{[\mu} v^{\nu]} \} \ , \ 
    \frac{\bnslash}{2}\gamma_\perp^\alpha \gamma^{[ \mu} n^{\nu ] }
        - 2  g^{\alpha [ \mu}_{\bot} \gamma^{\nu ]}\ ,\ 
    \frac{\bnslash}{2}\gamma_\perp^\alpha n^{[ \mu} v^{\nu ]} 
        + 2 g^{\alpha [ \mu}_{\bot} v^{\nu ]}
     \Big\} \,,\nn\\[4pt]
%
%\Upsilon^{\alpha \mu \nu}\sub{c1-6} &= \Big\{ \frac{1}{2} \Nbs
%\gammapa \Gamma^{\mu \nu}\sub{1-4}\ ,\
%    g^{\alpha [ \mu}_{\bot} \gamma^{\nu ]}\ ,\
%    g^{\alpha [ \mu}_{\bot} v^{\nu ]} \Big\}
%  \,,\nn \\
%
\Upsilon^{\alpha \beta \mu \nu}\sub{d1-6}
    &= \Big\{ \gpab \Gamma^{\mu \nu}\sub{1-4} \ ,\
    \frac{1}{4} \Nbs \gamma^{\{ \alpha}_{\bot} 
     g^{\beta \} [ \mu}_{\bot} \gamma^{\nu ] }\ ,\
    \frac{1}{4} \Nbs \gamma^{\{ \alpha}_{\bot} 
    g^{\beta \} [ \mu}_{\bot} v^{\nu ] } \Big\} 
   \,, \nonumber\\[4pt]
\Upsilon^{\alpha \beta \mu \nu}\sub{e1-10}
    &=\Big\{ 2\gpab \Gamma^{\mu\nu}\sub{1-4}\ , \
    \gammapa \gammapb \Gamma^{\mu\nu}\sub{1-4}\ , \
    \half \Nbs \gammapb g^{\alpha[\mu}_{\bot} \gamma^{\nu]}\ , \
    \half \Nbs \gammapb g^{\alpha[\mu}_{\bot} v^{\nu]}   \Big\} 
    \,, \nonumber \\[4pt]
\Upsilon^{\alpha \beta \mu \nu}\sub{f,h,i1-10}
    &=\Big\{ 2\gpab \Gamma^{\mu\nu}\sub{1-4}\ , \
    \gammapa \gammapb \Gamma^{\mu\nu}\sub{1-4}\ , \
    %\half \Nbs 
    \gammapb g^{\alpha[\mu}_{\bot} \gamma^{\nu]}\ , \
    %\half \Nbs
    \gammapb g^{\alpha[\mu}_{\bot} v^{\nu]}   \Big\} 
    \,, 
\end{align}
where $g_T^{\alpha\beta} = g^{\alpha\beta} - v^\alpha v^\beta$. 
Similarly, for the tensor four quark operator currents, a complete basis is
\begin{align} \label{Tensor4}
\big(\Upsilon \ot \Upsilon & \big)_{(j1-10)}^{\mu\nu}
    =\big(\Upsilon \ot \Upsilon \big)_{(k1-10)}^{\mu\nu}
   \nn\\
   &=\Big\{
    \Nbs\ot\Gamma^{\mu\nu}_{(1-4)} \,,\ 
    \Nbs \gamma^5\ot \gamma^5 \Gamma^{\mu\nu}_{(1-4)}
     \,,\  i\sigma^{\mu\nu} \ot 1\,,\
     i\sigma^{\mu\nu}\gamma_5 \ot \gamma_5
    \Big\} \,,
\end{align}
where just as for the vector case we have made use of chirality.

The relations for tensor Wilson coefficients obtained by solving the RPI-\$
constraint equation are:
\begin{align} \label{Tensorstart}
 A_{a1}^{(t)}(\omega)&=C_{1}(\omega)  \,,
 & A_{a2}^{(t)}(\omega)&=C_{2}(\omega) \,,\\
A_{a3}^{(t)}{(\omega)}&=C_{3}(\omega) \,,
& A_{a4}^{(t)}(\omega)&= C_{4}(\omega) 
 \,,\nn\\
A_{a5}^{(t)}(\omega) &= 2 \omega C'_1(\omega) \,, 
&A_{a6}^{(t)}(\omega) &= 2 \omega C'_2(\omega)
  \,,\nn\\
A_{a7}^{(t)}(\omega) &=-2 C_3(\omega) +2 \omega C'_3(\omega) \,,
&A_{a8}^{(t)}(\omega) &=-2 C_4(\omega) +2 \omega C'_4(\omega) \,. \nn % \\
\end{align}
The relations for Wilson coefficients from the RPI-a constraint equations are
\begin{align} 
 A_{b1-4}^{(t)}(\omega) &= \omega C_{1-4}^{(t)\prime}(\omega) ,
 & A_{c1-4}^{(t)}(\omega) &= C_{1-4}^{(t)}(\omega) .
\end{align}
Finally, solving the RPI-$\star$ constraint in Eq.~(\ref{constraint2}) for the
tensor case gives
\begin{align}
A_{d1}^{(t)}{(\omega)}&=-\omega C_1^{(t)\prime}(\omega) \,,
&A_{d2}^{(t)}(\omega)&=-\omega C_2^{(t)\prime}(\omega)-2C_{3}^{(t)}(\omega)
   \,,\nn\\
A_{d3}^{(t)}(\omega)&=-\omega C_3^{(t)\prime}(\omega)+2C_{3}^{(t)}(\omega) 
   \,,
&A_{d4}^{(t)}(\omega)&=-\omega C_4^{(t)\prime}(\omega) + 2C_{4}^{(t)}(\omega) 
   \,,\nn\\
A_{d5}^{(t)}(\omega)&=-2C_{3}^{(t)}(\omega) \,,
&A_{d6}^{(t)}(\omega)&=2C_{4}^{(t)}(\omega) \,.
\end{align}
while the constraint in Eq.~(\ref{constraint3}) has the solution 
\begin{align} \label{Tensorend}
A_{e1}^{(t)}(\omega_{1,2}) 
&=- \Big(\ffrac{m}{\omega_2}\Big)C_{1}^{(t)}(\omega_1\plus\omega_2)
   -2B_{b3}^{(t)}(\omega_{1,2})
   \,, \nonumber\\[4pt]
A_{e2}^{(t)}(\omega_{1,2}) 
&=-\Big(\ffrac{m}{\omega_2}\Big)C_{2}^{(t)}(\omega_1\plus\omega_2)
 %  -B_{b3}^{(t)}(\omega_{1,2}) %[?]
   -B_{b4}^{(t)}(\omega_{1,2})+B_{b5}(\omega_{1,2})
   \,, \nonumber\\[4pt]
A_{e3}^{(t)}(\omega_{1,2})
&=-\Big(\ffrac{m}{\omega_2}\Big)C_{3}^{(t)}(\omega_1\plus\omega_2) 
 -B_{b3}^{(t)}(\omega_{1,2})% [?]
 -B_{b5}^{(t)}(\omega_{1,2})
\,, \nonumber\\[4pt]
A_{e4}^{(t)}(\omega_{1,2}) 
&=-\Big(\ffrac{m}{\omega_2}\Big)C_{4}^{(t)}(\omega_1\plus\omega_2) 
 -2B_{b3}^{(t)}(\omega_{1,2})+B_{b4}^{(t)}(\omega_{1,2})
 +B_{b6}^{(t)}(\omega_{1,2})
\,, \nonumber\\[4pt]
A_{e5}^{(t)}(\omega_{1,2})
&=-\Big(\ffrac{m}{\omega_1\plus\omega_2}\Big)C_{1}^{(t)}(\omega_1\plus\omega_2) 
  -B_{b1}^{(t)}(\omega_{1,2})+4B_{b3}^{(t)}(\omega_{1,2})
\,, \nonumber\\[4pt]
A_{e6}^{(t)}(\omega_{1,2})
&=-\Big(\ffrac{m}{\omega_1\plus\omega_2}\Big)C_{2}^{(t)}(\omega_1\plus\omega_2) 
 -2B_{b1}^{(t)}(\omega_{1,2})+B_{b2}^{(t)}(\omega_{1,2})
 +2B_{b4}^{(t)}(\omega_{1,2})
\,, \nonumber\\[4pt]
A_{e7}^{(t)}(\omega_{1,2})
&=-\Big(\ffrac{m}{\omega_1\plus\omega_2}\Big)C_{3}^{(t)}(\omega_1\plus\omega_2) 
+3B_{b3}^{(t)}(\omega_{1,2})
\,, \nonumber\\[4pt]
A_{e8}^{(t)}(\omega_{1,2})
&=-\Big(\ffrac{m}{\omega_1\plus\omega_2}\Big)C_{4}^{(t)}(\omega_1\plus\omega_2) 
+6B_{b3}^{(t)}(\omega_{1,2})-3B_{b4}^{(t)}(\omega_{1,2})
\,, \nonumber\\[4pt]
A_{e9}^{(t)}(\omega_{1,2})
&=-2B_{b3}^{(t)}(\omega_{1,2})+B_{b5}^{(t)}(\omega_{1,2})
\,, \nonumber\\[4pt]
A_{e10}^{(t)}(\omega_{1,2})
&= 4B_{b3}^{(t)}(\omega_{1,2})
   -2B_{b4}^{(t)}(\omega_{1,2})+B_{b6}^{(t)}(\omega_{1,2}) \,.
\end{align}
The following Wilson coefficients of the ${\mathcal O}(\lambda^2)$ tensor currents
are not determined by the RPI constraints
\begin{align} \label{Tensornot}
 & A_{f1-10}^{(t)}(\omega_{1,2}) \,, & 
 & A_{g1-4}^{(t)}(\omega_{1,2,3}) \,, &
 & A_{h,i1-10}^{(t)}(\omega_{1,2}) \,, &
 & A_{j,k1-10}^{(t)}(\omega_{1,2,3}) \,.
\end{align}

%%%%%%%%%%%%%%%%%%%%%%%%%%%%%%%%%%%%%%%%%%%%%%%%%%%%%%%%%%%%%%%%%%%%%%%%

\subsection{Absence of supplementary projected operators at ${\cal O}(\lambda^2)$}
\label{sect_constD}

Here we show that the analysis above on the surface $v_\perp=0$ is complete by
showing that there are no supplementary projected operators as defined in
section~\ref{sect_rpi}. The analysis of the proceeding section makes this
simpler, since a complete set of relations have been derived for all currents
$J_j^{(2b-2e)}$.  Thus, we only need to worry about supplementary projected
operators generated by transforming the currents $J_j^{(2a,2f-2k)}$. To simplify
our proof we first swap all factors of $\bn\mcdot v$ for $1/(n\mcdot v)$.

First consider the SCET RPI-II transformation at ${\cal O}(\lambda^0)$ for these
$J^{(2)}$ currents. At this
order we have
\begin{align}
 &  \bn^\mu \to \bn^\mu +\epsilon_\perp^\mu \,,\qquad
 & \gamma_\perp^\mu \to  \gamma_\perp^\mu - \frac{n^\mu}{2}\,
  \epsilon_\perp\!\!\!\!\!\!\slash\: - \frac{\epsilon_\perp^\mu}{2} &\, \nslash
   \,,
 & v_\perp^\mu \to - \frac{n^\mu}{2}\, \epsilon_\perp\mcdot v_\perp 
   -  \frac{\epsilon_\perp^\mu}{2}\, n\mcdot v \,,
  \nn\\
 &  {\cal P}_\perp^\mu \to {\cal P}_\perp^\mu - \frac{n^\mu}{2}\, \epsilon_\perp
  \mcdot {\cal P}_\perp \,,\qquad
 &  {\cal B}_\perp^\mu \to  {\cal B}_\perp^\mu - \frac{n^\mu}{2}\, \epsilon_\perp
  \mcdot  {\cal B}_\perp & \,.\qquad
\end{align}
We use the convention where all indices $\alpha\beta$ are $\perp$ for the field
structures and Dirac structures in $J_j^{(2f-2k)}$.  Now due to the contractions
of the $\alpha$ and $\beta$ indices only the transformations on $\bn^\mu$ and
$\gamma_\perp^\mu$ can contribute for these operators (there are no $\bn$'s or
$\perp$'s in the $J^{(2a)}$ case). The transformation related to their labels
$\omega_i$ is ${\cal O}(\lambda)$ and need not be considered and the field
transformations cancel. Thus, the only terms that appear in an RPI-II relation
are those whose Dirac structure transforms, $\delta_{\rm II}^\lambda
\Upsilon_{(a,f,g,h,i,j,k)}\ne 0$. However, with our choice of the complete basis
of Dirac structures on the $v_\perp=0$ surface, the structures for these
currents all have zero transformations. In this regard it was important to take a
basis with no factors of $\bnslash$. Away from this surface we must add to our
basis of Dirac structure by including additional $v_\perp$ dependent terms and
it is only these terms that can have additional relations. For example, factors
of $v_\perp$ are induced when we reduce a basis that includes factors of
$\bnslash$ using the trace formula in Eq.~(\ref{reduce1}). The same is true with
our choice of the basis of $J_j^{(1b)}$ currents.

Finally consider whether the transformations RPI-$\star$ and RPI-\$ induce SPO's
or equivalently SCET RPI-I and HQET RPI. Since the $\bn$ transformation in
RPI-\$ did not enter at the order we are working it is apparent that there are
no SPO's from the HQET RPI. Examining the results of the RPI-I transformations
we find that none of the $J_j^{(2a,2f-2k)}$ currents have ${\cal O}(\lambda^0)$
transformations (since the Dirac structures transform at ${\cal O}(\lambda)$ and
the field structures that do transform all cancel out).

Thus the results derived in the previous section give the complete set of RPI
relations for the ${\cal O}(\lambda^2)$ currents when $v_\perp=0$.

%%%%%%%%%%%%%%%%%%%%%%%%
%%%%%%%%%%%%%%%%%%%%%%%%
\section{Change of Basis and Comparison with Tree Level Results}
\label{sect_tree}
%%%%%%%%%%%%%%%%%%%%%%%%
%%%%%%%%%%%%%%%%%%%%%%%%

In expanding the heavy-to-light currents, two different bases of operators are
useful. At tree level it is convenient to write the result for the currents in
terms of collinear covariant derivatives, giving one basis. For the derivation
of RPI relations and factorization theorems, a basis such as the one in
Eq.~(\ref{JNNLO}) is more useful.

The tree level matching of the full theory current $\bar {q}
\overline{\Gamma} {b}$ onto SCET currents was done to subsubleading order in
Ref.~\cite{Beneke:2002ph}. In deriving Feynman rules we find the momentum space
version more convenient so we use the equivalent result from~\cite{Lee:2004ja}
\begin{eqnarray}\label{Jfulltree}
\bar\psi^q \overline{\Gamma}\psi^b \rightarrow
\overline{J}^{(0)}\plus \overline J^{(1a)}\plus\overline J^{(1b)}\plus\overline J^{(2a)}\plus\overline J^{(2b)}\plus
\overline J^{(2c)}\plus\overline J^{(2d)}\plus\overline J^{(2d)}\plus\overline J^{(2f)}\,,
\end{eqnarray}
where
\begin{align} \label{J1}
  \overline J^{(0)}(\omega) &=
   \bar\chi_{n,\omega} \overline \Gamma {\mathcal H}_v \,, \\[4pt] %\qquad
 \overline J^{(1a)}(\omega) &=
    \frac{1}{\omega} \big( \bar\chi_n\, i\!\! \cDgppPl
    \big)_{\omega}
    \overline \Theta_{(a) \: \alpha} \: {\mathcal H}_v \,, %\nn\\
 &\overline J^{(1b)}(\omega_{1,2})
    &=  \frac{-1}{m} \: \bar\chi_{n,\omega_1}
    (ig {\mathcal B}_{c\perp}^\alpha)_{\omega_2}\,
    \overline \Theta_{(b) \, \alpha}
    \, {\mathcal H}_v \,.  \nn\\[4pt]
  \overline J^{(2a)}(\omega) &=  \frac{1}{2m}\: \bar\chi_{n,\omega} \,
  \overline \Upsilon_{(a) \: \alpha}
    {\cDpusRight{\alpha} } \, {\mathcal H}_v
   \,, %\nn\\
&\overline J^{(2b)}(\omega) &= \frac{1}{\omega}\:
\bar\chi_{n,\omega} \,
   \overline \Upsilon_{(b)\:\alpha}  \,
    \cDpusLeft{\alpha}\, {\mathcal H}_v \,, \nn\\[4pt]
\overline J^{(2c)}(\omega) &= - \Big(\bar\chi_n
\frac{\bn\mcdot v}{\bnP}\,
  \frac{ ig n\mcdot {\mathcal B}_c}{n\mcdot v} \Big)_{\omega}
   \overline \Upsilon_{(c)}\: {\mathcal H}_v
    \,, %\nn\\
 &\overline J^{(2d)}(\omega_{1,2}) &=  \frac{-1}{m}
   \bar\chi_{n,\omega_1}
   \Big(\frac{ig n\mcdot {\mathcal B}_c}{n\mcdot v} \Big)_{\omega_2}
   \overline \Upsilon_{(d)}\,  {\mathcal H}_v  \,,
   \nn\\[4pt]
\overline J^{(2e)}(\omega_{1,2}) &= \frac{-1}{m\,\omega_1}
   \big( \bar\chi_n i\! \cDgppPl \big)_{\omega_1}
   ( ig {\mathcal B}_{c\perp}^\beta)_{\omega_2} \overline \Upsilon_{(e) \: \alpha\beta}\:
   {\mathcal H}_v 
   \,, \hspace{-1.5cm} \nn\\[4pt]
\overline J^{(2f)}(\omega_{1,2}) &=  \frac{-1}{m\,}
   \bar\chi_{n,\omega_1}
    \Big(\frac{1}{ n\mcdot v\,\bnP}  i {\mathcal D}_{c\perp}^{\alpha}
   \: ig {\mathcal B}_{c\perp}^{\beta} \Big)_{\omega_2}
   \overline \Upsilon_{(f) \: \alpha\beta}\:
   {\mathcal H}_v 
   \,. \nn \hspace{-1.5cm}
\end{align}
The $\overline{\Gamma}$ in $\overline{J}^{(0)}$ is simply the Dirac structure of
the full theory current. The Dirac structures that appear in the subleading
currents are
\begin{align}
\overline \Theta_{(a) \: \alpha}   &=
    \gamma_\alpha^\perp \frac{\bnslash}{2 \nb \cdot v} \overline \Gamma
     \,, %\nn \\
 &\overline \Theta_{(b) \: \alpha} &=
    \overline \Gamma  \frac{\nslash}{2 n \cdot v} \gamma_\alpha^\perp
     \,
\end{align}
and
\begin{align}
 \overline \Upsilon_{(a) \: \alpha} &=
    \overline \Gamma  \gamma^T_\alpha
     \,, %\nn \\
  &\overline \Upsilon_{(b) \: \alpha} &=
     \gamma_\alpha^\perp \frac{\bnslash}{2 \nb \cdot v} \overline \Gamma
     \,, %\nn \\
  &\overline \Upsilon_{(c)} &=
    \overline \Gamma
     \,, \nn \\[5pt]
  \overline \Upsilon_{(d)} &=
    \overline \Gamma  \frac{\nslash\bnslash}{4}
     \,,  %\nn\\
  &\overline \Upsilon_{(e) \: \alpha \beta} &=
     \gamma^\perp_\alpha \frac{\bnslash}{2} \overline \Gamma \frac{\nslash}{2}
     \gamma^\perp_\beta
     \,, %\nn \\
  & \overline \Upsilon_{(f) \: \alpha \beta} &=
    \overline \Gamma \gamma^\perp_\alpha \gamma^\perp_\beta
  \,.
\end{align}
Each of the operators $\overline J$ has unit Wilson coefficient at tree level.
By re-expressing these operators in the basis of operators presented in this
paper, we determine the tree-level Wilson coefficients of our currents.  This
provides a check of the RPI relations.

%%%%%%%%%%%%%%%%%%%%%%%%%
\subsection{Conversion} 
%%%%%%%%%%%%%%%%%%%%%%%%%

In terms of our basis, the leading order tree level current
$\overline{J}^{(0)}$ is given by
\begin{eqnarray} \label{J0conv}
 \int\!\! \dd{\omega} \overline J^{(0)}(\omega) 
 = \int\!\!\dd{\omega}J^{(0)}_1(\omega)\,.
\end{eqnarray}
This result holds for all five Lorentz types, $\overline
\Gamma=\{1\,,\gamma^5\,,\gamma^\mu\,,\gamma^5\gamma^\mu,i\sigma^{\mu\nu}\}$.
For the remainder of this section, we will suppress the explicit
$\omega$-dependence of our basis $J(\omega)$'s as well as the appropriate
integrals $\int [\dd\omega_{i}]$ whenever results hold equally well as integrals
or as densities. For example, \Eqr{J0conv} would be written simply as $\overline
J^{(0)}=J^{(0)}_{1}$. If the Lorentz type ($s,p,v,a,t$) of the current is not
specified, the same result holds for all five types as above.

For the $\O(\lambda)$ currents, the relations differ for the scalar, vector, and tensor cases,
\begin{align}
 \int\!\! d\omega_1\: \overline J^{(1a)}_{s,p}(\omega_1) &= 
  \int\!\! d\omega_1\: J^{(1a)}_1(\omega_1)
  -\int\!\! d\omega_{1,2}\: \frac{m}{\omega_1\plus\omega_2}
   J^{(1b)}_1(\omega_1,\omega_2)
  \,, \hspace{-8cm}  \\[6pt]
 \int\!\! d\omega_1\: \overline J^{(1a)}_{v,a}(\omega_1) &= 
  \int\!\! d\omega_1\: J^{(1a)}_1(\omega_1)
  +\int\!\! d\omega_{1,2}\: \frac{m}{\omega_1\plus\omega_2}
   \big[J^{(1b)}_1(\omega_1,\omega_2)-2J^{(1b)}_2(\omega_1,\omega_2)\big]
  \,, \nn \\[6pt]
 \int\!\! d\omega_1\: \overline J^{(1a)}_{t}(\omega_1) &= 
  \int\!\! d\omega_1\: J^{(1a)}_1(\omega_1)
  +\int\!\! d\omega_{1,2}\: \frac{m}{\omega_1\plus\omega_2}
   \big[{-}J^{(1b)}_1(\omega_1,\omega_2)-2J^{(1b)}_2(\omega_1,\omega_2)\big]
  \,, \nn \\[6pt]
 \overline J^{(1b)}_{s,p} &= 0 \,, 
 \qquad \overline J^{(1b)}_{v,a} = -J^{(1b)}_3 \,, %\nn\\[4pt] 
 \qquad \overline J^{(1b)}_{t} = 2J^{(1b)}_1 + J^{(1b)}_3
   +2J^{(1b)}_4+2J^{(1b)}_5
  \,. \nn
\end{align}
The last line of relations are true as integrals or as densities. For example
$J^{(1b)}_{v,a}(\omega_{1,2}) = -J^{(1b)}_3(\omega_{1,2})$. At $\O(\lambda^2)$
the relations between the two forms of subleading currents are the same for all
$\overline J^{(2a,2b,2c)}$ currents
\begin{align}
\overline J^{(2a)} &= J^{(2a)}_1
  \,, %\nn\\
&\overline J^{(2b)} &=J^{(2c)}_1
  \,, %\nn\\[4pt] 
&\int\!\! d\omega_1\: \overline J^{(2c)}& 
  = \int\!\! d\omega_{1,2}\:
   \frac{m}{\omega_2}\Big[{-}J^{(2e)}_1 \!+\! J^{(2g)}_1\Big]
   \,,
\end{align}
where in the last relation the arguments of $\overline J^{(2c)}(\omega_1)$ and
$J^{(2e,2g)}_1(\omega_{1,2})$ are implicit.
The remaining currents come in different combinations depending on the Dirac
structure. For $\overline J^{(2d)}$ we have
\begin{align}
 \overline J^{(2d)}_{s,p} &= \, 0
  \,, %\nn\\[4pt] 
&\overline J^{(2d)}_{v,a} &=
  J^{(2e)}_3 \!-\!  J^{(2g)}_3
  \,, %\nn\\[4pt]
& \overline J^{(2d)}_{t} &= {-}J^{(2e)}_3
  \!+\! J^{(2g)}_3
 \,.
\end{align}
For the scalar $\overline J^{(2e)}$ currents,
\begin{eqnarray}
\int\!\! d\omega_{1,2}\: \overline J^{(2e)}_{s,p}
\!\! &=& \!\!\!
   \int\!\! d\omega_{1,2}\:\Big[J^{(2e)}_1 \!-\! J^{(2e)}_2\Big] 
   \!+\! \int\!\! d\omega_{1,2} \Big[e\to f \Big]  
   \!+\! \int\!\! d\omega_{1,2,3}\: 
  \frac{\omega_2\plus\omega_3}{\omega_1\plus\omega_2}\Big[e\to h \Big] 
   \,, %\\[4pt]
\end{eqnarray}
with similar relations for the vector and tensor cases (suppressing the
integrals for convenience),
\begin{eqnarray}
\overline J^{(2e)}_{v,a}
&=&\Big[
    J^{(2e)}_1-J^{(2e)}_3 \!-\! J^{(2e)}_4 \!+\! J^{(2e)}_6 \!+\! 2J^{(2e)}_7
    \Big]
    +  \Big[
    J^{(2f)}_1-J^{(2f)}_3 \!-\! J^{(2f)}_4 \!+\! J^{(2f)}_6 \!-\! 2J^{(2f)}_7
    \Big] \nonumber \\
  %\nn\\[4pt]
%& & \qquad
  && 
  +\frac{\omega_2\plus\omega_3}{\omega_1\plus\omega_2}  \Big[
  J^{(2h)}_1-J^{(2h)}_3 \!-\! J^{(2h)}_4 \!+\! J^{(2h)}_6 \!-\! 2J^{(2h)}_7
  \Big] 
  \,,\nn\\[4pt]
\overline J^{(2e)}_{t}
&=& \Big[
    -J^{(2e)}_1-J^{(2e)}_3Á+J^{(2e)}_5+J^{(2e)}_7
    \Big] 
 % \nn\\[4pt] 
 % & & \qquad 
 + \Big[e\to f\Big]
 +\frac{\omega_2\plus\omega_3}{\omega_1\plus\omega_2}\Big[e\to h \Big]\,.
\end{eqnarray}
Finally for $\overline J^{(2f)}$, 
\begin{eqnarray}\label{J2fconv}
\overline J^{(2f)}_{s,p}
&=& \Big[{-}J^{(2e)}_1+J^{(2e)}_2\Big]
-\frac{\omega_1}{\omega_2}\Big[e\to f\Big] -\Big[e\to h \Big]
 \,,\\[4pt]
\overline J^{(2f)}_{v,a}
&=&  \Big[{-}J^{(2e)}_3-J^{(2e)}_4+2J^{(2e)}_6\Big] 
-\frac{\omega_1}{\omega_2}\Big[e\to f\Big] -\Big[e\to h \Big]
\,,\nn\\[4pt]
\overline J^{(2f)}_{t}
&=& \Big[{-}J^{(2e)}_1-J^{(2e)}_3+J^{(2e)}_5+2J^{(2e)}_7\Big]  
 -\frac{\omega_1}{\omega_2} \Big[e\to f\Big] -\Big[e\to h \Big]\,,
 \nn
\end{eqnarray}
where the suppressed integrals are the same as for $\overline J^{(2e)}$.

%%%%%%%%%%%%%%%%%%%%%%%%%%%%%%%%%%%%%%%%%%%%%%%%%%%%%%%%%%%%%%%%%%%%%
\subsection{Wilson coefficients at tree level} \label{sect_tree2}
%%%%%%%%%%%%%%%%%%%%%%%%%%%%%%%%%%%%%%%%%%%%%%%%%%%%%%%%%%%%%%%%%%%%%

Inserting Eqs. (\ref{J0conv}-\ref{J2fconv}) into \Eqr{Jfulltree}, we can read
off the tree level Wilson coefficients of our basis.  For example, since $\overline{J}^{(0)}$
is the only term at leading order we have $C^{(d)}_{1}(\omega)=1$ and
$C^{(d)}_{j\neq1}(\omega)=0$ for $d=s,p,v,a,t$.

%\subsubsection{Scalars and Pseduoscalars}

For scalar currents, the non-vanishing tree-level Wilson coefficients 
are 
\begin{align}
C^{(s)}_1(\omega_{})&= 1\,,
& B^{(s)}_{a1}(\omega_{})&=1\,,
& B^{(s)}_{b1}(\omega_{1,2})&=\frac{-m}{\omega_1\plus\omega_2}
\,,
\end{align}
and at ${\mathcal O}(\lambda^2)$
\begin{align} 
 &A^{(s)}_{a1}(\omega_{})=1 \,,
 & &A^{(s)}_{c1}(\omega_{})=1
   \,,\nn\\ 
 &A^{(s)}_{e1}(\omega_{1,2})=-\frac{m}{\omega_2}
  \,,
 & &A^{(s)}_{f1}(\omega_{1,2})=1+\frac{\omega_1}{\omega_2}
   \,,\nn\\
 &A^{(s)}_{f2}(\omega_{1,2})=-1 - \frac{\omega_1}{\omega_2} \,,
 &&A^{(s)}_{g1}(\omega_{1,2})=\frac{m}{\omega_2}
   \,,\nn\\[4pt]
 &A^{(s)}_{h1}(\omega_{1,2})=\frac{\omega_1\plus2\omega_2\plus\omega_3}
   {\omega_1\plus\omega_2} \,,
 &&A^{(s)}_{h2}(\omega_{1,2,3})=-\frac{\omega_1\plus2\omega_2\plus\omega_3}
   {\omega_1\plus\omega_2}\,.
\end{align}
The same results hold for the pseudoscalar currents.
%\subsubsection{Vectors and Axial-Vectors}
To $\O(\lambda^2)$, the values of the Wilson coefficients for vector currents
that do not vanish at tree-level are
\begin{align}
C^{(v)}_1(&\omega_{})= 1
\,,& B^{(v)}_{a1}(&\omega_{})=1
\,,\nn\\ B^{(v)}_{b1}(&\omega_{1,2})=\frac{m}{\omega_1\plus\omega_2}
   \,,
& B^{(v)}_{b2}(&\omega_{1,2})= \frac{-2m}{\omega_1\plus\omega_2} \,,
& B^{(v)}_{b3}(&\omega_{1,2})= -1\,,
\end{align}
and
\begin{align}
 &A^{(v)}_{a1}(\omega_{})=1 \,,
 & &A^{(v)}_{c1}(\omega_{})=1
   \,,\nn\\[5pt] 
 &A^{(v)}_{e1}(\omega_{1,2})=1-\frac{m}{\omega_2}
  \,,
 & &A^{(v)}_{e3}(\omega_{1,2})=-1
  \,,\nn\\ 
 &A^{(v)}_{e4}(\omega_{1,2})=-2
  \,,
 & &A^{(v)}_{e6}(\omega_{1,2})=3
  \,,\nn\\ 
 &A^{(v)}_{e7}(\omega_{1,2})= 2
   \,,
 & &A^{(v)}_{f1}(\omega_{1,2})=1   \,,\nn\\ 
 &A^{(v)}_{f3}(\omega_{1,2})=-1+\frac{\omega_1}{\omega_2}
    \,,
 & &A^{(v)}_{f4}(\omega_{1,2})=-1+\frac{\omega_1}{\omega_2}
   \,,\nn\\[4pt] 
 &A^{(v)}_{f6}(\omega_{1,2})= 1-\frac{2\omega_1}{\omega_2} \,,
 & &A^{(v)}_{f7}(\omega_{1,2})= -2
   \,,\nn\\[4pt] 
 &A^{(v)}_{g1}(\omega_{1,2})=\frac{m}{\omega_2} 
   \,,
 & &A^{(v)}_{g3}(\omega_{1,2})=-1
    \,,\nn\\[6pt] 
 &A^{(v)}_{h1}(\omega_{1,2,3})=\frac{\omega_2\plus\omega_3}{\omega_1\plus\omega_2}
   \,,
 & &A^{(v)}_{h3}(\omega_{1,2,3})=\frac{\omega_1\minus\omega_3}
  {\omega_1\plus\omega_2}
  \,,\nn\\[4pt]
 &A^{(v)}_{h4}(\omega_{1,2,3})=\frac{{\omega_1}\minus\omega_3}
   {\omega_1\plus\omega_2}
   \,,
 & &A^{(v)}_{h6}(\omega_{1,2,3})=\frac{{-2\omega_1}\minus\omega_2\plus\omega_3}
    {\omega_1\plus\omega_2}
   \,,\nn\\[4pt]
 &A^{(v)}_{h7}(\omega_{1,2,3})=-2\:\frac{\omega_2\plus\omega_3}
   {\omega_1\plus\omega_2}
   \,.
\end{align}
The same results hold for the axial vector currents.
%\subsubsection{Tensors}
Finally, for the ${\mathcal O}(\lambda^2)$ tensor currents we have nonvanishing
coefficients
\begin{align}
 C^{(t)}_1(\omega_{}) &= 1 \,, 
 &B^{(t)}_{a1}(\omega_{}) &=1\,,
 &B^{(t)}_{b1}(\omega_{1,2}) &=2 - \frac{m}{\omega_1\plus\omega_2} \,,  
 &B^{(t)}_{b2}(\omega_{1,2})=& \frac{-2m}{\omega_1\plus\omega_2} 
    \,,\nn\\[4pt]
  B^{(t)}_{b3}(\omega_{1,2}) &=1 \,,
 & B^{(t)}_{b4}(\omega_{1,2}) &=2  \,,
 & B^{(t)}_{b5}(\omega_{1,2}) &=2  \,,
\end{align}
and
\begin{align}
 &A^{(t)}_{a1}(\omega_{})=1
  \,,
 & &A^{(t)}_{c1}(\omega_{})=1
  \,,\nn\\ 
 &A^{(t)}_{e1}(\omega_{1,2})=-2-\frac{m}{\omega_2}
  \,,
 & &A^{(t)}_{e3}(\omega_{1,2})=-3
   \,,\nn\\ 
 &A^{(t)}_{e5}(\omega_{1,2})=2
   \,,
 & &A^{(t)}_{e7}(\omega_{1,2})=3
   \,,\nn\\[4pt] 
 &A^{(t)}_{f1}(\omega_{1,2})
   =-1+\frac{\omega_1}{\omega_2}
   \,,
 &&A^{(t)}_{f3}(\omega_{1,2})=-1+\frac{\omega_1}{\omega_2}
  \,,\nn\\[4pt]
 &A^{(t)}_{f5}(\omega_{1,2})=1-\frac{\omega_1}{\omega_2}
  \,,
 & &A^{(t)}_{f7}(\omega_{1,2})=1-\frac{2\omega_1}{\omega_2}
  \,,\nn\\[4pt] 
 &A^{(t)}_{g1}(\omega_{1,2})=\frac{m}{\omega_2}
   \,,
 & &A^{(t)}_{g3}(\omega_{1,2})=1
   \,,\nn\\[4pt]
 &A^{(t)}_{h1}(\omega_{1,2,3})=\frac{\omega_1\minus\omega_3}
   {\omega_1\plus\omega_2}
   \,, 
 & &A^{(t)}_{h3}(\omega_{1,2,3})=\frac{\omega_1\minus\omega_3}
   {\omega_1\plus\omega_2}
   \,,\nn\\[4pt]
 &A^{(t)}_{h5}(\omega_{1,2,3})=\frac{-\omega_1\plus\omega_3}
   {\omega_1\plus\omega_2}
   \,,
 & &A^{(t)}_{h7}(\omega_{1,2,3})=\frac{{-2}\omega_1\minus\omega_2\plus\omega_3}
   {\omega_1\plus\omega_2}\,.
\end{align}
It is straightforward to check that these results all satisfy the RPI relations
from section~\ref{sect_constC}, providing a cross-check on those results.

%%%%%%%%%%%%%%%%%%%%%%%%

\subsection{One-Loop Results} \label{sect_one}

The relations from section~\ref{sect_constC} apply at any order in perturbation
theory, so they can also be used to determine one-loop values for certain
coefficients. For the LO currents the one-loop coefficients in $\overline {\rm
  MS}$ at $\mu=m$ are~\cite{Bauer:2000yr}
\begin{eqnarray} \label{matchC}
C_{1}^{(s)}(\hat\omega)
 &=& 1 - \frac{\alpha_s(m)C_F}{4\pi} \bigg\{ 2 \ln^2 (\hat\omega)
  + 2 {\textrm{Li}}_2(1\!-\!\hat\omega) - \frac{2\ln(\hat\omega)}{1-\hat\omega} +\frac{\pi^2}{12}
  \bigg\} \,, \nn \\*
C_{1}^{(v)}(\hat\omega)
 &=& 1 - \frac{\alpha_s(m)C_F}{4\pi} \bigg\{ 2\!\ln^2(\hat\omega)
  + 2 {\textrm{Li}}_2(1\!-\!\hat\omega)
  + \ln(\hat\omega) \Big( \frac{3\hat\omega-2}{1-\hat\omega}\Big)
  + \frac{\pi^2}{12} + 6\bigg\} , \nn \\*
C_{1}^{(t)}(\hat\omega)
 &=& 1 - \frac{\alpha_s(m)C_F}{4\pi} \bigg\{ 2\!\ln^2(\hat\omega)
  + 2 {\textrm{Li}}_2(1\!-\!\hat\omega)
  +  \ln(\hat\omega) \Big( \frac{4\hat\omega-2}{1-\hat\omega} \Big)
  + \frac{\pi^2}{12} + 6 \bigg\} , \nn \\
C_{2}^{(v)} (\hat\omega,1)
 &=& \frac{\alpha_s(m)C_F}{4\pi}\:  \bigg\{ \frac{2}{(1-\hat\omega)}
  + \frac{2\hat\omega\ln(\hat\omega)} {(1-\hat\omega)^2} \bigg\} \,, 
  \qquad %\nn \\
C_{2}^{(t)}(\hat\omega)
 = 0  \,, \nn \\
C_{3}^{(v)}(\hat\omega)
 &=& \frac{\alpha_s(m)C_F}{4\pi} \bigg\{
  \frac{(1-2\hat\omega)\hat\omega \ln(\hat\omega) }{(1-\hat\omega)^2}
  - \frac{\hat\omega}{1-\hat\omega} \bigg\}
 \,, \nn \\*
C_{3}^{(t)}(\hat\omega)
 &=& \frac{\alpha_s(m)C_F}{4\pi} \bigg\{
    \frac{-2\hat\omega \ln(\hat\omega)}{1-\hat\omega} \bigg\}
 \,, \qquad
C_{4}^{(t)}(\hat\omega)
 = 0 \,,
\end{eqnarray}
where $\hat\omega=\omega/m$ and $C_F=4/3$ for color $SU(3)$.  The
quark-gluon-antiquark operators $J^{(1b)}$ have coefficients that are not fixed
by RPI, and these were determined by a one-loop matching
in~\cite{Beneke:2004rc,Becher:2004kk}. Thus all ${\mathcal O}(\lambda^{0,1})$
currents are known at one-loop order. The expressions are fairly lengthy, and so
we do not repeat them here. Using their results and our
Eqs.~(\ref{Scalarstart}-\ref{Scalarend}), (\ref{Vectorstart}-\ref{Vectorend}),
and (\ref{Tensorstart}-\ref{Tensorend}), the coefficients of the currents
$J^{(2a,2b,2c,2d,2e)}$ are also determined at one-loop order.

We give the scalar current case as an example.  For the
scalar current, the coefficient at $\mu=m$ is~\cite{Beneke:2004rc,Becher:2004kk} 
 \begin{eqnarray} \label{Bb1}
B_{(b1)}^{(s)}(\hat\omega_{1,2},1) &=& -\frac{1}{\hat\omega} 
 +\frac{\alpha_s C_F}{4\pi} \bigg[ 
  \frac{-2}{\hat\omega_2} \bigg\{ \ln^2 \hat\omega \!-\! \ln^2 \hat\omega_1 \!-\!
   \ln \Big ( \frac{\hat\omega}{\hat\omega_1} \Big) \bigg\}
   + \,\Big(\frac{4}{\hat\omega} \!+\!
   \frac{2}{1\!-\!\hat\omega} \Big) \ln \hat\omega
   \nonumber \\
&&  - \frac{2}{\hat\omega_1}\bigg (\frac{\ln
   \hat\omega}{1\!-\!\hat\omega}-\frac{\ln \hat\omega_2}{1\!-\!\hat\omega_2} 
   \bigg ) -\frac{\hat\omega_2\ln \hat\omega_2}{(1\!-\!\hat\omega_2)^2}
   \!+\! \frac{2(1\!-\!\hat\omega_1)}{\hat\omega_1 \hat\omega_2}\big\{
{\textrm{Li}}_2 (1\!-\!\hat\omega) \!-\! {\textrm{Li}}_2 (1\!-\!\hat\omega_1) \big\}
    \nonumber \\
&& 
   \nonumber \\
&& - \frac{2}{\hat\omega_1 \hat\omega_2} \Big\{ {\textrm{Li}}_2
(1-\hat\omega_2)-\frac{\pi^2}{6}\Big\}
-\,\frac{4}{\hat\omega}-\frac{1}{1-\hat\omega_2}\bigg]
\nonumber \\[5pt]
 && - \frac{\alpha_s C_A}{4\pi} \bigg[ 
   \frac{-1}{\hat\omega_2} \Big\{ \ln^2 \hat\omega \!-\! \ln^2 \hat\omega_1 
   \!-\!  \ln \Big( \frac{\hat\omega}{\hat\omega_1} \Big)\Big\}
   + \frac{1}{\hat\omega_1} \ln \Big ( \frac{\hat\omega}{\hat\omega_2} \Big)
   + \frac{\ln \hat\omega_2}{1\!-\! \hat\omega_2} \nonumber \\
&& \,  +
\frac{1-\hat\omega_1}{\hat\omega_1 \hat\omega_2}\big\{ {\textrm{Li}}_2
(1\!-\!\hat\omega) - {\textrm{Li}}_2 (1\!-\!\hat\omega_1) \big\} 
 -\frac{1}{\hat\omega_1 \hat\omega_2} \Big\{ {\textrm{Li}}_2
(1\!-\!\hat\omega_2)-\frac{\pi^2}{6}\Big\} \bigg] \nn\\[4pt]
&&
+\frac{\alpha_s C_F}{4\pi}\frac{1}{\hat\omega} \bigg[2 \ln^2(\hat\omega)
  + 2 {\rm Li}_2(1\!-\!\hat\omega) - \frac{2\ln(\hat\omega)}{1-\hat\omega}
  +\frac{\pi^2}{12} \bigg]
\,,
\end{eqnarray}
where $C_A = 3$, $\hat \omega_{1,2}=\omega_{1,2}/m$,
$\hat\omega=\hat\omega_1+\hat\omega_2$, and we have transformed to our basis. We
will also need the derivative of the LO scalar currents coefficient
\begin{eqnarray}\label{C1p}
 \frac{d}{d \omega} C_1^{(s)}(\hat\omega, 1) &=& -\frac{\alpha_s
 (m) C_F}{2 \pi m}\left\{ \frac{-1 + \hat\omega + (2 - 4 \hat\omega
+\hat\omega^2) \ln \hat\omega}{\hat\omega (1-\hat\omega)^2} \right\} \,.
\end{eqnarray}
Now in section IIIC we derived the following results for the ${\mathcal
  O}(\lambda^2)$ currents
\begin{align}
 A_{a1}^{(s)}(\omega)&=C_{1}^{(s)}(\omega) \,, &
 A_{a2}^{(s)}(\omega)&=2 \omega C_1^{(s)\,\prime}(\omega)\,, \\[4pt]
 A_{b1}^{(s)}(\omega)& = \omega  C_{1}^{(s)\,\prime}(\omega) \,,
 & A_{c1}^{(s)}(\omega) &= C_{1}^{(s)}(\omega)\,,\quad
 A_{d1}^{(s)}(\omega) =-\omega C_1^{(s)\,\prime}(\omega) \,, \nn \\[4pt]
 A_{e1}^{(s)}(\omega_{1,2})
  &=-\ffrac{m}{\omega_2}\,
 C_1^{(s)}(\omega_{1}\!+\!\omega_2) \,, %\nn \\[4pt]
 & A_{e2}^{(s)}(\omega_{1,2})&=
  -\fbrac{m}{\omega_1\! +\! \omega_2}\, C_{1}^{(s)}(\omega_1\! +\! \omega_2)
  - B_{b1}^{(s)}(\omega_{1,2}) \,. \nn
\end{align}
Combined with Eqs.(\ref{matchC}-\ref{C1p}), these relations determine the
coefficients at one-loop order. The results for the $J^{(2a-2e)}_j$ vector and
  tensor currents at one-loop order are easily obtained in the same manner.

%%%%%%%%%%%%%%%%%%%%%%%%

\section{Conclusion}

In this paper we derived a complete basis of scalar, vector, and tensor
heavy-to-light currents at next-to-next-to-leading order in the power counting,
${\mathcal O}(\lambda^2)$.  Building on the approach in Ref.~\cite{Hill:2004if}
where one takes $v_\perp=0$ from the start, we constructed the full set of RPI
relations that leave us on this surface. The completeness of deriving RPI
relations projected on a surface was analyzed.  With a careful choice of Dirac
structures in our analysis of heavy-to-light currents at ${\mathcal
  O}(\lambda^2)$ it was demonstrated that the projected RPI gives the full set
of constraints.  We also investigated the path dependence of Wilson lines in
order to clarify what conditions they must obey to give the correct cancellation
of usoft gluon effects, and to demonstrate the manner in which results are
independent of the choice of boundary condition.

A simple method for counting the number of Dirac structures in the basis for any
operator with $d=4$ was given.\footnote{We did not consider the complication
  that occurs if one uses dimensional regularization where there can be additional
  evanescent ${\cal O}(\lambda^2)$ operators which vanish for $d=4$.  In SCET
  this type of operator has been studied for the ${\cal O}(\lambda)$ currents in
  Ref.~\cite{Becher:2004kk}.}  Several types of reparameterization invariance
provide restrictions on the structure of these currents, which we formulated as
constraint equations on the allowed Dirac structures and Wilson coefficients as
given in Eqs.~(\ref{constraint1}), (\ref{constrainth}), (\ref{constrainta1}),
(\ref{constrainta2}), (\ref{constraint2}), and (\ref{constraint3}). We expect
that a similar setup with constraint equations and projected surfaces will be
useful in deriving RPI relations at higher orders in $\lambda$ and in deriving
results for non heavy-to-light currents.

Our main results are contained in the solution of the constraint equations as
given in Eqs.~(\ref{Scalar}-\ref{Scalarend}), (\ref{Vector}-\ref{Vectorend}),
and (\ref{Tensor}-\ref{Tensorend}). These results determine the coefficients of
five of the eleven NNLO operators, $J^{(2a,2b,2c,2d,2e)}_j$, for various Dirac
structures indicated by $j$ and at any order in perturbation theory, in terms of
the coefficients of NLO and LO operators. This determines $7$, $23$, and $32$
Wilson coefficients for the scalar, vector, and tensor heavy-to-light currents
respectively. Results at tree-level and one-loop order were discussed in
sections~\ref{sect_tree2} and \ref{sect_one}. Finally, the operators
$J^{(2f,2g,2h,2i,2j,2k)}$ defined in Eqs.~(\ref{JNNLO},\ref{JNNLO4}) together
with the Dirac structures in
Eqs.~(\ref{Scalar},\ref{Scalar4},\ref{Vector},\ref{Vector4},\ref{Tensor},\ref{Tensor4})
were shown to not be constrained by reparameterization invariance.

We thank D.Pirjol for useful comments, and the Institute of Nuclear Theory for
their hospitality while parts of this work were completed. This work was
supported by the Office of Nuclear Science and U.S.\ Department of Energy under
DE-FG02-93ER-40762 (J.K.)  and the cooperative research agreement
DF-FC02-94ER40818 (C.A. and I.S.), as well as the DOE OJI program and Sloan
Foundation (I.S.).

\bibliography{aks}

\end{document}